\documentclass[12pt]{article}
\usepackage{amssymb}
\usepackage{epsfig}
\begin{document}
\title{Spectral Properties of the $k$--Body Embedded Gaussian 
Ensembles of Random Matrices \\} 

\author{
\\ \\
L. Benet, T. Rupp and H. A. Weidenm\"uller\\
Max--Planck--Institut f\"ur Kernphysik \\
D--69029 Heidelberg, Germany\\ \\ \\}
\date{\today}

\maketitle

\begin{abstract}
We consider $m$ spinless Fermions in $l > m$ degenerate
single--particle levels interacting via a $k$--body random interaction
with Gaussian probability distribution and $k \leq m$ in the limit $l
\rightarrow \infty$ (the embedded $k$--body random ensembles). We
address the cases of orthogonal and unitary symmetry. We derive a
novel eigenvalue expansion for the second moment of the Hilbert--space
matrix elements of these ensembles. Using properties of the expansion
and the supersymmetry technique, we show that for $2k > m$, the
average spectrum has the shape of a semicircle, and the spectral
fluctuations are of Wigner--Dyson type. Using a generalization of the
binary correlation approximation, we show that for $k \ll m \ll l$,
the spectral fluctuations are Poissonian. This is consistent with the
case $k = 1$ which can be solved explicitly. We construct limiting
ensembles which are either fully integrable or fully chaotic and show
that the $k$--body random ensembles lie between these two extremes.
Combining all these results we find that the spectral correlations for
the embedded ensembles gradually change from Wigner--Dyson for $2k >
m$ to Poissonian for $k \ll m \ll l$.

\end{abstract}

\newpage

\section{Introduction}
\label{int}

The stochastic behavior displayed by spectra and wave functions of
quantum many--body systems (atoms, molecules, atomic nuclei, quantum
dots) is usually modelled in terms of canonical random--matrix theory
(RMT); the application of RMT to such systems has been very successful
in many cases \cite{bro81,guh98}. At the same time, this type of
modelling is not completely realistic: All many--body systems that
occur in nature are effectively governed by one-- and two--body
forces, while the use of canonical RMT is tantamount to assuming
many--body forces between the constituents. Indeed, assuming that the
one-- and two--body interaction can be modelled stochastically, one
finds that the resulting number of independent random variables is
much smaller than in RMT. For instance, the number of independent
two--body matrix elements in a shell--model calculation in atoms or
nuclei is typically much smaller than the dimension $N$ of the
matrices involved, while the number of independent random variables in
RMT is of order $N^2$. This poses the question whether a more
realistic stochastic modelling of many--body systems might yield
results which differ from RMT predictions. The question was addressed
in the 1970's with the help of numerical simulations using matrices of
fairly small dimensions. The main results were: In a certain limit,
the average level density does not have the shape of a semicircle but
is Gaussian; the ensembles are neither stationary nor ergodic;
unfolding of the spectra yields Wigner--Dyson spectral fluctuation
properties, see Refs.~\cite{fre71} to \cite{bro76}. Interest in model
Hamiltonians with random two--body interactions has resurged in recent
years in several areas of many--body physics (see Refs.~\cite{flam96}
to \cite{kot00}), and the question of possible further differences
between such models and RMT has resurfaced.

The paradigmatic models for this type of question are the $k$--body
embedded Gaussian random ensembles EGOE($k$) and EGUE($k$) introduced
by Mon and French in 1975 \cite{mon75} and fully defined below. The
integer $k$ takes any value between $1$ and $m$, the total number of
particles. A number of useful and interesting results for these
ensembles have been obtained \cite{mon75,bro81,ver84}. In particular,
it was shown that in the dilute limit (number $m$ of Fermions small in
comparison with the number $l$ of single--particle states) and for $k
\ll m$, the shape of the average spectrum is Gaussian (rather than a
semicircle as in the case of RMT). However, a comprehensive analytical
approach to the spectral fluctuation properties of these ensembles is
still lacking. To quote from Ref.~\cite{bro81}: ``... there is no real
theory yet for EGOE fluctuations, the gap being one of the most
significant ones in the entire subject''. As far as we know, the
situation has not changed in the last twenty years. The gap is due to
the fact that, in contrast to RMT, the embedded ensembles do not
possess the (unitary, orthogonal or symplectic) invariance in Hilbert
space which is so essential for the successful analytical treatment of
RMT. 

In this paper, we report on a novel approach to the $k$--body
embedded Gaussian ensembles. Although physical interest focusses on
the case $k = 2$, the generic features are better understood by
addressing the case of arbitrary $k$ with $1 \leq k \leq m$. As always
in RMT, we consider the limit of infinite matrix dimension, realized
by letting $l \rightarrow \infty$. One essential tool in our analysis
is provided by the orthogonal or unitary invariance of EGOE($k$) or
EGUE($k$) with respect to the single--particle states from which
Hilbert space is constructed. We show that this invariance takes over
the role of the (orthogonal or unitary) invariance of standard RMT. We
first focus attention on the second moment $A^{(k)}$ of the random
$k$--body interaction in the Hilbert space of many--body states. We
introduce a novel eigenvector expansion for $A^{(k)}$. This allows us
to derive statements about the shape of the average spectrum as a
function of $k$ and $m$. Further information on the spectral behavior
of EGOE($k$) and EGUE($k$) is obtained with the help of the
supersymmetry method~\cite{efe83,ver85} which applies thanks to the
eigenvector expansion of $A^{(k)}$. We derive and solve the
saddle--point equations for EGOE($k$) and EGUE($k$). We show that the
mean level density and the spectral fluctuations coincide with those
of canonical RMT. Most importantly, we calculate the first
non--vanishing term in the loop expansion and thereby estimate the
range of validity of the saddle--point approximation. This range is
given by $2k > m$. This result is consistent with an estimate obtained
by calculating the second and fourth moments of the interaction. In
order to gain access to the regime $2k < m$, we generalize the
``binary correlation'' method of Mon and French~\cite{mon75} to the
calculation of level correlation functions. The method applies in the
limit $k \ll m \ll l$. We show that in this limit, the spectral
fluctuations are Poissonian. This is consistent with explicit results
for $k = 1$. To reach a physical understanding of our results, we
construct limiting ensembles of random matrices which are either fully
integrable or fully chaotic. We show that EGOE($k$) and EGUE($k$) lie
between these two extremes. In this way, we show that -- contrary to
common belief -- for $2k \lesssim m$ and $l \rightarrow \infty$,
EGOE($k$) and EGUE($k$) do not follow Wigner--Dyson statistics, and we
give a simple argument why this is so. We comment on the apparent
agreement between canonical RMT and data from numerical simulation and
from atomic and nuclear physics and argue that this agreement is
caused by a finite--size effect which disappears in the limit $l
\rightarrow \infty$.

The embedded ensembles and the limiting ensembles are defined in
Section~\ref{def}. The second moment $A^{(k)}$ is introduced, and an
important identity derived, in Section~\ref{bas}. The eigenvector
expansion derived in Section~\ref{eig} is a central analytical result
and enables us to calculate the low moments of the interaction
(Section~\ref{low}) and to apply the supersymmetry method
(Section~\ref{sup}). The case $k \ll m \ll l$ is studied in
Section~\ref{bin} while Section~\ref{con} contains the conclusions. A
preliminary account of some of the results has appeared in
Refs.~\cite{wei00} and~\cite{ben00}.

\section{Definitions and Elementary Facts}
\label{def}

\subsection{Definitions}
\label{defa}

We consider $m$ spinless Fermions in $l > m$ degenerate
single--particle states with associated creation and annihilation
operators $a_j^{\dagger}$ and $a_j$ where $j = 1,\ldots,l$. The ratio
$f = m/l$ is called the filling factor. For reasons explained in
Section~\ref{par}, we confine ourselves to the case of less than half
filling, $f \leq 1/2$ or $2m \leq l$. Hilbert space is spanned by
orthonormal vectors labelled $| \mu \rangle, | \nu \rangle, \ldots$.
Each such vector is given by a Slater determinant $a_{j_1}^{\dagger}
\ldots a_{j_m}^{\dagger} | 0 \rangle$ where $| 0 \rangle$ denotes the
vacuum. Uniqueness is guaranteed by the requirement $j_1 < j_2 <
\ldots < j_m$. The dimension $N$ of this space is
\begin{equation}
\label{eq2.1}
N = { l \choose m } \ .
\end{equation}
The random $k$--body interaction $V_k$ with $k \leq m$ has the form
\begin{equation}
\label{eq2.2}
V_k \ = \sum_{{1 \leq j_1 < j_2 < \ldots < j_k \leq l} \atop {1 \leq
  i_1 < i_2 < \ldots < i_k \leq l}} v_{j_1,\ldots,j_k; i_1,\ldots,i_k}
  a_{j_1}^{\dagger} \ldots a_{j_k}^{\dagger} a_{i_k} \ldots a_{i_1} \ .
\end{equation}
We refer to $k$ as to the rank of the interaction. As in the canonical
case, we use the labels $\beta = 1$ and $\beta = 2$ for the orthogonal
and the unitary ensemble, respectively. For $\beta = 1$, the matrix
elements are real and obey
\begin{equation}
\label{eq2.3b}
v_{j^{}_1,\ldots,j^{}_k; i^{}_1,\ldots,i^{}_k} =
  v^*_{j^{}_1,\ldots,j^{}_k; i^{}_1,\ldots,i^{}_k} =
  v_{i^{}_1,\ldots,i^{}_k; j^{}_1, \ldots, j^{}_k} \ \ (\beta = 1) \ .
\end{equation}
For $\beta = 2$, the matrix elements are complex and obey 
\begin{equation}
\label{eq2.3a}
v_{j_1,\ldots,j_k; i_1,\ldots,i_k} = v^*_{i_1,\ldots,i_k;
  j_1,\ldots,j_k} \ \ (\beta = 2) \ .
\end{equation}
Matrix elements not linked by the symmetry relations~(\ref{eq2.3b}) or
(\ref{eq2.3a}) are uncorrelated Gaussian distributed random variables
with zero mean and a common second moment denoted by $v_0^2$. Thus,
\begin{eqnarray}
\label{eq2.4}
\overline{v_{j^{}_1,\ldots,j^{}_k; i^{}_1,\ldots,i^{}_k}
    v_{j'_1,\ldots,j'_k; i'_1,\ldots,i'_k}} &=& v_0^2 [\delta_{j^{}_1
    i'_1} \ldots \delta_{j^{}_k i'_k} \delta_{i^{}_1 j'_1} \ldots
    \delta_{i^{}_k j'_k} \nonumber \\
&&\qquad + \delta_{\beta 1} \delta_{j^{}_1 j'_1} \ldots
    \delta_{j^{}_k j'_k} \delta_{i^{}_1 i'_1} \ldots \delta_{i^{}_k
    i'_k}]  \ .
\end{eqnarray}
The overbar denotes the average over the ensemble.

We ``embed'' the $k$--body interaction $V_k(\beta)$ in the
$m$--particle space by taking matrix elements $\langle \nu | V_k | \mu
\rangle$. For $\beta = 1 \ (\beta = 2)$, this defines the $k$--body
embedded Gaussian orthogonal (unitary) ensemble of random matrices,
respectively, in short EGOE($k$) and EGUE($k$). We note that for $k = m$,
we have done no more than rewrite the canonical ensembles GOE and GUE
in what may appear a complicated way. Naturally, interest focusses on
the case $k < m$. However, it is always useful to check that for $k =
m$ our results coincide with canonical random--matrix theory.

Eqs.~(\ref{eq2.3b}),(\ref{eq2.3a}) and (\ref{eq2.4}) are invariant
under arbitrary orthogonal ($\beta = 1$) or unitary ($\beta = 2$)
transformations of the single--particle states and so are, therefore,
the ensembles EGOE($k$) and EGUE($k$).

The choice of $v_0^2$ determines the energy scale of our problem.
Without loss of generality, we put $v_0^2 = 1$ in the sequel.

\subsection{Regular Graphs}
\label{reg1}

A graphical representation of the ensemble is obtained by assigning to
each Hilbert space vector $| \mu \rangle$ a vertex $\mu$, and to each
non--diagonal matrix element $\langle \nu | V_k | \mu \rangle$ which
is not identically equal to zero, a link connecting the vertices $\mu$ 
and $\nu$. The diagonal matrix elements
$\langle \mu | V_k | \mu \rangle$ are represented by loops attached to
the vertices $\mu $. It is instructive to explore the
properties of the resulting topological structure.

The number of vertices is obviously given by $N$, the dimension of
Hilbert space. The number $M$ of links emanating from any given vertex
is the same for all vertices and is given by
\begin{equation}
\label{eq2.5}
M = \sum_{s = 1}^k {m \choose s} {l-m \choose s} \ . 
\end{equation}
To see this, we observe that for the non--diagonal matrix element
$\langle \nu | V_k | \mu \rangle$ not to vanish identically, the
vectors $| \mu \rangle$ and $| \nu \rangle$ may differ in the
occupation of as many as $s$ single--particle states where $s =
1,2,\ldots,k$. The number $M$ is obtained by summing over all these
possibilities. With the exception of the loops, the resulting
structure is what is called a ``regular graph'' in the mathematical
literature~\cite{bol85}. When we represent the structure as a matrix,
the number of non--diagonal matrix elements which do not vanish
identically is equal in each row and each column and given by $M$.
For $k < m$, we have $M < N-1$ while $M = N-1$ for $k = m$. It is also
of interest to calculate the total number $P$ of independent links.
This number is given by the number of matrix elements above the main
diagonal which do not vanish identically. Hence,
\begin{equation}
\label{eq2.5a}
P = \frac{1}{2} M N.
\end{equation}
The minimum number $L_{\rm min}$ of links needed to get from an
arbitrary initial vertex $\mu $ to an arbitrary final vertex
$\nu$ is bounded by 
\begin{equation}
\label{eq2.6}
L_{\rm min} \leq  \biggl[ \frac{m}{k} \biggr] \ .
\end{equation}
The bracket denotes the largest integer smaller than or equal to $m/k$.
The inequality~(\ref{eq2.6}) holds because the maximum number of
particles that must be moved, is $m$. The maximum number of particles
that can be moved per link is $k$. For $k = m$ we get $L_{\rm min} =
1$. We observe that the graph is very densely connected: For $l \gg 1$,
$L_{\rm min}$ is much smaller than $N$. Obviously, $N$, $M$, $P$ and
$L_{\rm min}$ do not depend on $\beta$. 

\begin{figure}
\begin{minipage}{15cm}
\centerline{\psfig{figure=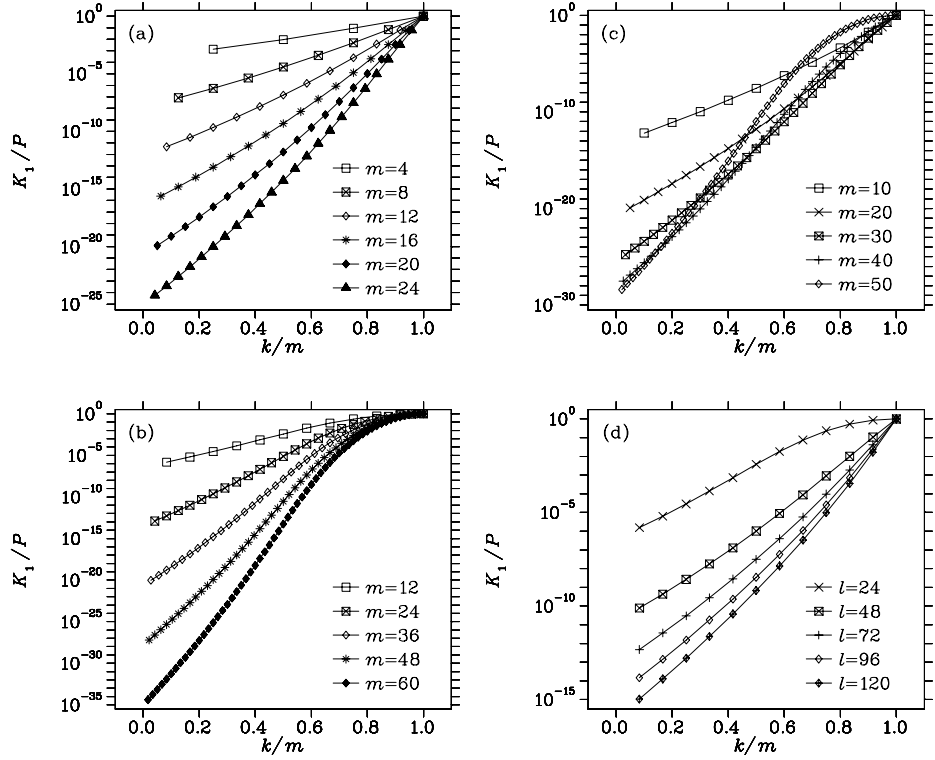,width=10cm,angle=90}}
\end{minipage}
\caption{The ratio $K_1/P$, see Eqs.~(\ref{eq2.5a}) and (\ref{eq2.7}),
  on a logarithmic scale versus $k/m$. Panel (a): $f = m/l = 1/5$.
  Panel (b): $f = m/l = 1/2$. Panel (c): $l = 100$. Panel (d): $m = 12$.}
\label{fig1}
\end{figure}

The number $K_{\beta}$ of independent random variables is different
for the unitary and the orthogonal ensemble and essentially given by
the square of the number of independent choices of the indices
$j_1,\ldots,j_k$ subject to the constraint $j_1 < j_2 < \ldots <
j_k$. More precisely,
\begin{equation}
\label{eq2.7}
K_{\beta} = \frac{\beta}{2} {l \choose k} \biggl[{l \choose k} +
\delta_{\beta 1} \biggr] \ .
\end{equation}
The ratio $K_{\beta}/P$ of the number $K_{\beta}$ of independent
random variables and the number $P$ of different links is, for $k \ll
m$, much smaller than one and, for $f \leq 1/2$, approaches the value
$\beta (N + \delta_{\beta 1})/(N - 1)$ monotonically from below as $k$
approaches $m$. This shows that for $k \ll m$ there are strong
correlations between matrix elements on different links. The
correlations disappear as we approach the canonical limit $k = m$.
Fig.~\ref{fig1} shows $\log_{10} K_{1}/P$ versus $k/m$ for various
values of the parameters $l$ and $m$.

\subsection{Limiting Ensembles}
\label{lim}

Fig.~\ref{fig1} suggests that EGOE($k$) and EGUE($k$) both lie between
two limits, obtained by assigning the minimum and maximum number,
respectively, of independent random variables to the graph structure
defined by the operators appearing in Eq.~(\ref{eq2.2}). It is helpful
to study the properties of these limiting ensembles. The minimum
number of random variables is obviously one. The corresponding
ensemble EGE$_{\rm min}(k$) is given by the matrix elements $\langle
\nu | V_k^{\rm min} | \mu \rangle$ of the interaction
\begin{equation}
\label{eq2.8}
V_k^{\rm min} = v \sum_{{1 \leq j_1 < j_2 < \ldots < j_k \leq l} \atop
  {1 \leq i_1 < i_2 < \ldots < i_k \leq l}} a_{j_1}^{\dagger} \ldots
  a_{j_k}^{\dagger} a_{i_k} \ldots a_{i_1} \ .
\end{equation}
The factor $v$ is a Gaussian random variable and real (complex) for
$\beta = 1$ ($\beta = 2$, respectively). Without loss of generality
we may, however, put $\overline{|v|^2} = 1$, removing the distinction
between the unitary and the orthogonal cases. The ensembles EGOE$_{\rm
  max}(k$) and EGUE$_{\rm max}(k$) containing the maximum number of
independent random variables are obtained by assigning (within the
constraints imposed by symmetry) a different random variable $v_{\nu
  \mu}$ to each link in the graph. The interaction has the matrix
elements
\begin{equation}
\label{eq2.9}
v_{\nu \mu} \sum_{{1 \leq j_1 < j_2 < \ldots < j_k
  \leq l} \atop {1 \leq i_1 < i_2 < \ldots < i_k \leq l}} \langle \nu |
  a_{j_1}^{\dagger} \ldots a_{j_k}^{\dagger} a_{i_k} \ldots a_{i_1} |
  \mu \rangle \ .
\end{equation}
For $\beta = 1 \ (\beta = 2)$, the matrix $v_{\nu \mu}$ is real
symmetric (complex Hermitean, respectively). Elements not connected by
symmetry are uncorrelated Gaussian random variables with mean value
zero and variance $\overline{v_{\nu \mu} v_{\nu' \mu'}} = \delta_{\mu
  \nu'} \delta_{\nu \mu'} + \delta_{\beta 1} \delta_{\mu \mu'}
\delta_{\nu \nu'}$. 

The ensemble EGE$_{\rm min}(k$) is fully integrable and has spectral
fluctuations which are not of Wigner--Dyson type. We demonstrate this
for the two extreme cases $k = 1$ and $k = m$. For $k = 1$, the
interaction has the form $( \sum_{j = 1}^l a_j^{\dagger} ) ( \sum_{i =
  1}^l a_i )$ which obviously commutes with all permutations of the
labels $j = 1,\ldots,l$ and $i = 1,\ldots,l$ of the single--particle
states. Hence, the eigenvectors of the interaction are simultaneously
eigenvectors of irreducible representations of the symmetric group:
The system is integrable, and the spectral fluctuations are not
Wigner--Dyson. The Gaussian distribution of the factor $v$ in
Eq.~(\ref{eq2.8}) does not affect this statement as it only smears
the entire spectrum of a given representation of the ensemble over an
interval of width unity. We have not succeeded in diagonalizing the
interaction matrix analytically for $v = 1$. However, extensive
numerical calculations have consistently yielded two different
degenerate eigenvalues $\lambda_1, \lambda_2$ which are reproduced by
the formulas $\lambda_1 = l$ and $\lambda_2 = 0 $, with degeneracies
$n_1 = {l-1 \choose m-1}$ and $n_2 = {l-1 \choose m}$.

For $k = m$, it is easy to see that the matrix representation of
EGE$_{\rm min}(m$) in Hilbert space carries the entry $v$ on every
element. Diagonalization of this matrix is trivial and gives the
eigenvalues $N v$ (non--degenerate) and zero ($(N-1)$--fold
degenerate). Again, the ensemble is fully integrable, and the spectral
fluctuations are not Wigner--Dyson.

We show in Section~\ref{sup} that the spectral fluctuation properties
of the ensembles EGOE$_{\rm max}(k$) and EGUE$_{\rm max}(k$) coincide
with those of the GOE and GUE, respectively. (In actual fact, we do so
only for EGUE$_{\rm max}(k$)). Given this fact, we see that the two
limiting ensembles cover the extreme cases of fully integrable and
fully chaotic systems, respectively. Fig.~\ref{fig1} then suggests
that as $k$ increases, the spectral fluctuations of both EGOE($k$) and
EGUE($k$) may undergo a gradual transition from Poissonian to
Wigner--Dyson behavior. In what follows, we present conclusive evidence
for this statement.
 
\section{The Second Moment of the Interaction}
\label{bas}

After these preliminaries, we turn to the central object of the
theory: The second moment of the matrix $\langle \nu | V_k(\beta) |
\mu \rangle$. For $\beta = 2$ we derive an identity for this quantity
which is very helpful for what follows. 

\subsection{The Second Moment}
\label{sec}

By virtue of the randomness of $V_k(\beta)$, the matrix $\langle \nu |
V_k(\beta) | \mu \rangle$ is a random matrix with a Gaussian
probability distribution and zero mean value. The distribution is
completely specified in terms of the second moment $A^{(k)}_{\mu \nu,
  \rho \sigma}(\beta)$ defined by
\begin{equation}
\label{eq3.1}
A^{(k)}_{\mu \nu, \rho \sigma}(\beta) = \overline{\langle \mu | V_k(\beta)
  | \sigma \rangle \langle \rho | V_k(\beta) | \nu \rangle} \ .
\end{equation}
Using Eqs.~(\ref{eq2.2}),(\ref{eq2.4}) and $v^2 = 1$, we find
\begin{eqnarray}
\label{eq3.2}
A^{(k)}_{\mu \nu, \rho \sigma}(\beta) &=& \sum_{{1 \leq j_1 < j_2 <
  \ldots < j_k \leq l} \atop {1 \leq i_1 < i_2 < \ldots < i_k \leq l}}
  \biggl[ \langle \mu |a_{j_1}^{\dagger} \ldots a_{j_k}^{\dagger}
  a_{i_k} \ldots a_{i_1} | \sigma \rangle \nonumber \\
&&\qquad\times [ \langle \rho | a_{i_1}^{\dagger} \ldots
  a_{i_k}^{\dagger} a_{j_k} \ldots a_{j_1} | \nu \rangle \nonumber \\
&&\qquad + \delta_{\beta 1} \langle \rho | a_{j_1}^{\dagger} \ldots
  a_{j_k}^{\dagger} a_{i_k} \ldots a_{i_1} | \nu \rangle ] \biggr] \ .
\end{eqnarray}
The matrix elements appearing on the right--hand side of
Eq.~(\ref{eq3.2}) are manifestly real. Using this fact and the
identity $\langle \rho | H | \nu \rangle^* = \langle \nu | H^{\dagger}
| \rho \rangle$ valid for any operator $H$, we can rewrite
Eq.~(\ref{eq3.2}) in the form
\begin{eqnarray}
\label{eq3.3}
A^{(k)}_{\mu \nu, \rho \sigma}(\beta) &=& \sum_{{1 \leq j_1 < j_2 <
  \ldots < j_k \leq l} \atop {1 \leq i_1 < i_2 < \ldots < i_k \leq l}}
  \biggl[ \langle \mu |a_{j_1}^{\dagger} \ldots a_{j_k}^{\dagger}
  a_{i_k} \ldots a_{i_1} | \sigma \rangle \nonumber \\
&&\qquad\times [ \langle \rho | a_{i_1}^{\dagger} \ldots
  a_{i_k}^{\dagger} a_{j_k} \ldots a_{j_1} | \nu \rangle \nonumber \\
&&\qquad + \delta_{\beta 1} \langle \nu | a_{i_1}^{\dagger} \ldots
  a_{i_k}^{\dagger} a_{j_k} \ldots a_{j_1} | \rho \rangle ] \biggr] \
  .
\end{eqnarray}
Eq.~(\ref{eq3.3}) is the central relation for the theory developed
in this paper. As a check, we consider Eq.~(\ref{eq3.3}) for $k = m$
and $\beta = 2$. There is only one term in the sum over the indices
$i_{\alpha}, \alpha = 1, \ldots, m$ for which $a_{i_m} \ldots a_{i_1}
| \sigma \rangle$ gives a non--zero result. This is the term by which
all the $m$ Fermions in $| \sigma \rangle$ are annihilated. The same
combination of indices also appears in the expression $\langle \rho |
a_{i_1}^{\dagger} \ldots a_{i_m}^{\dagger}$. For this expression not
to vanish, we must have $\sigma = \rho$. Likewise, we conclude that
$A^{(k)} = 0$ unless $\mu = \nu$, and similarly for the last term
($\beta = 1$) in Eq.~(\ref{eq3.3}). As a result, we find
\begin{equation}
\label{eq3.4}
A^{(m)}_{\mu \nu, \rho \sigma}(\beta) =  \delta_{\mu \nu} \delta_{\rho
  \sigma} + \delta_{\beta 1} \delta_{\mu \rho} \delta_{\nu \sigma} \ .
\end{equation}
This is the result of canonical random--matrix theory, usually written
in the form $ ( \lambda^2 / N ) (\delta_{\mu \nu} \delta_{\rho \sigma}
+ \delta_{\beta 2} \delta_{\mu \rho} \delta_{\nu \sigma})$ where $2
\lambda$ gives the radius of the semicircle. We note that with our
normalization we have $\lambda = \sqrt{N}$.

\subsection{Duality}
\label{dua}

The duality relation is very helpful later on in obtaining explicit
expressions for several quantities involving traces of powers of
$V_k(\beta)$, and in the application of the supersymmetry formalism.
We consider the case $\beta = 2$. For arbitrary $k \leq m$, the
duality relation has the form
\begin{equation}
\label{eq3.5}
A^{(k)}_{\mu \nu, \rho \sigma}(2) = A^{(m-k)}_{\mu \sigma, \rho
  \nu}(2) \ .
\end{equation}
This relation connects the matrix elements of the $k$--body
interaction with those of the $(m-k)$--body interaction. The crucial
difference is in the sequence of state vectors labelled $\mu, \nu,
\rho, \sigma$. The duality relation is obtained from a generalization
of the argument leading, for $k = m$, from Eq.~(\ref{eq3.3}) to
Eq.~(\ref{eq3.4}). We consider a fixed term in the sum over the $j$'s
and $i$'s in Eq.~(\ref{eq3.3}). For $A^{(k)}_{\mu \nu, \rho \sigma}(2)$
not to vanish, the single--particle states labelled $j_1,\ldots,j_k$
must be occupied in both $| \mu \rangle$ and $| \nu \rangle$, and the
states labelled $i_1,\ldots,i_k$ must be occupied in both $| \rho
\rangle$ and $| \sigma \rangle$. Moreover, $| \mu \rangle$ and $|
\sigma \rangle$ must have $(m-k)$ occupied single--particle states in
common, and the same holds for $| \rho \rangle$ and $| \nu \rangle$.
Hence, for each pair of sets of labels $j_{1},\ldots,j_k$ and
$i_{1}, \ldots,i_k$ for which $A^{(k)}(2)$ does not vanish, there
exists a pair of sets of $(m-k)$ uniquely defined labels
$p_{1} < \ldots <p_{m-k}$ and $q_{1} < \ldots < q_{m-k}$ such that the
four states labelled $\mu, \sigma, \rho, \nu$ have the form
\begin{eqnarray}
\label{eq3.6}
&&| \mu \rangle = (-)^{f_{\mu}} a_{j_1}^{\dagger} \ldots
a_{j_k}^{\dagger} a_{p_1}^{\dagger} \ldots a_{p_{m-k}}^{\dagger} | 0
\rangle \ ,
\nonumber \\
&&| \sigma \rangle = (-)^{f_{\sigma}} a_{i_1}^{\dagger} \ldots
a_{i_k}^{\dagger} a_{p_1}^{\dagger} \ldots a_{p_{m-k}}^{\dagger} | 0
\rangle \ ,
\nonumber \\
&&| \rho \rangle = (-)^{f_{\rho}} a_{i_1}^{\dagger} \ldots
a_{i_k}^{\dagger} a_{q_1}^{\dagger} \ldots a_{q_{m-k}}^{\dagger} | 0
\rangle \ ,
\nonumber \\
&&| \nu \rangle = (-)^{f_{\nu}} a_{j_1}^{\dagger} \ldots
a_{j_k}^{\dagger} a_{q_1}^{\dagger} \ldots a_{q_{m-k}}^{\dagger} | 0
\rangle \ .
\end{eqnarray}
The sets of indices \{$p_{\alpha}$\} and \{$q_{\alpha}$\} must have
zero overlap with the sets \{$i_{\alpha}$\} and \{$j_{\alpha}$\}. The
phase factors in Eq.~(\ref{eq3.6}) are not known and will not be needed.
Eq.~(\ref{eq3.6}) implies
\begin{eqnarray}
\label{eq3.7}
&&\langle \mu | a_{r_1}^{\dagger} \ldots a_{r_{m-k}}^{\dagger}
a_{t_{m-k}} \ldots a_{t_{1}} | \nu \rangle = (-)^{f_{\mu} + f_{\nu}} \
, \nonumber \\
&&\langle \rho | a_{t_1}^{\dagger} \ldots a_{t_{m-k}}^{\dagger}
a_{r_{m-k}} \ldots a_{r_{1}} | \sigma \rangle = (-)^{f_{\rho} +
  f_{\sigma}} \ , \nonumber \\
&&\langle \mu | a_{j_1}^{\dagger} \ldots a_{j_{k}}^{\dagger}
a_{i_{k}} \ldots a_{i_{1}} | \sigma \rangle = (-)^{f_{\mu} +
  f_{\sigma}} \ , \nonumber \\
&&\langle \rho | a_{i_1}^{\dagger} \ldots a_{i_{k}}^{\dagger}
a_{j_{k}} \ldots a_{j_{1}} | \nu \rangle = (-)^{f_{\rho} +
  f_{\nu}} \ .
\end{eqnarray}
From Eq.~(\ref{eq3.7}) we see that the product of the matrix elements
in the first two equations equals that of the matrix elements in the
last two equations. We sum the resulting equality over the sets
\{$i_{\alpha}$\} and \{$j_{\alpha}$\} and the sets \{$p_{\alpha}$\}
and \{$q_{\alpha}$\}, respectively, using the one--to--one
correspondence of both pairs of sets. This yields
\begin{eqnarray}
\label{eq3.8}
A^{(k)}_{\mu \nu, \rho \sigma}(2) &=& \sum_{{1 \leq j_1 < j_2 < \ldots
  < j_k \leq l} \atop {1 \leq i_1 < i_2 < \ldots < i_k \leq l}} \langle
  \mu |a_{j_1}^{\dagger} \ldots a_{j_k}^{\dagger} a_{i_k} \ldots
  a_{i_1} | \sigma \rangle \nonumber \\
&&\qquad\times \langle \rho | a_{i_1}^{\dagger} \ldots
  a_{i_k}^{\dagger} a_{j_k} \ldots a_{j_1} | \nu \rangle \nonumber \\
&=& \sum_{{1 \leq r_1 < r_2 < \ldots < r_{m-k} \leq l} \atop {1 \leq
  t_1 < t_2 < \ldots < t_{m-k} \leq l}} \langle \mu |a_{r_1}^{\dagger}
  \ldots a_{r_{m-k}}^{\dagger} a_{t_{m-k}} \ldots a_{t_1} | \nu \rangle
  \nonumber \\ 
&&\qquad\times \langle \rho | a_{t_1}^{\dagger} \ldots
  a_{t_{m-k}}^{\dagger} a_{r_{m-k}} \ldots a_{r_1} | \sigma \rangle \ .
\end{eqnarray}
This is Eq.~(\ref{eq3.5}).

\subsection{Particle--Hole Symmetry}
\label{par}

In shell--model calculations in atomic and nuclear physics,
particle--hole symmetry is often used to simplify the calculations
and to relate results for more than half--filled shells to those with
less than half filling. This procedure applies as well in the present
case but, as we shall see, is actually of little use because we
confine ourselves to random interactions with fixed rank $k$.

Let $| \tilde{0} \rangle = \prod_{j = 1}^l a_j^{\dagger} | 0 \rangle$
denote the ``full shell''. There exists a one--to--one correspondence
between the set of state vectors $| \mu \rangle$ containing $m$
particles defined in Section~\ref{defa}, and the set $| \tilde{\mu}
\rangle$ obtained by creating $(m-k)$ holes in the full shell,
$b_{i_1}^{\dagger} \ldots b_{i_{m-k}}^{\dagger} | \tilde{0} \rangle$.
The $b^{\dagger}$'s with $b_j^{\dagger} = a_j$ are the creation
operators for holes. We may consider the matrix elements of
$V_k(\beta)$ in the basis of hole states $| \tilde{\mu} \rangle$
rather than in that of the particle states $| \mu \rangle$. To make
this particle--hole transformation effective, we must rewrite the
interaction $V_k(\beta)$ in Eq.~(\ref{eq2.2}) in terms of the $b$'s
and $b^{\dagger}$'s and bring all hole creation operators up front.
The last step entails a number of commutators of $b$'s and
$b^{\dagger}$'s. As a result, we find that an interaction which has
rank $k$ in particle representation, will be a sum of interactions of
ranks $0, 1, \ldots, k$ in hole representation. This is why
particle--hole symmetry is not immediately useful in the present
context. On the other hand, the very use of a random interaction of
fixed rank $k$ is an idealization, and it is, therefore, legitimate to
consider such an interaction both for particles ($2m \leq l$) and for
holes ($2(l-m) \leq l$), the two types of interaction not being
connected by any symmetry relation. That is the point of view taken in
this paper. It is then obviously sufficient to study the particle
case, i.e., to confine $m$ to values $2m \leq l$ so that $f \leq 1/2$.

\section{Eigenvector Expansion of the Second \\ Moment}
\label{eig}

Progress in understanding the properties of EGOE($k$) and EGUE($k$)
rests on an eigenvalue expansion of the second moment $A^{(k)}(\beta)$.
To this end, we note that $A^{(k)}(\beta)$ defined in Eq.~(\ref{eq3.2})
has the symmetry properties $A^{(k)}_{\mu \nu, \rho \sigma}(2) =
A^{(k)}_{\rho \sigma, \mu \nu}(2)$ and $A^{(k)}_{\mu \nu, \rho
  \sigma}(2)^* = A^{(k)}_{\nu \mu, \sigma \rho}(2)$. Considered as a
``grand matrix'' carrying the indices $M = (\mu, \nu)$ and $M' = (\rho,
\sigma)$, respectively, the matrix $A^{(k)}_{\mu \nu, \rho \sigma}(2)$
is, therefore, Hermitean. (We use the term grand matrix to avoid
confusion with the term supermatrix in the supersymmetry approach).
Every Hermitean  matrix possesses a bilinear expansion in eigenvectors.
We construct this eigenvector expansion for $A^{(k)}_{M M'}(2)$. In
somewhat symbolic notation, the eigenvectors obey 
\begin{equation}
\label{eq3.9}
\sum_{M'} A^{(k)}_{M M'}(2) C^{(sa)}_{M'} = \Lambda^{(s)}(k)
C^{(sa)}_{M} \ .
\end{equation}
Here $s$ is a running label and the index $a$ allows for possible
degeneracies. When written explicitly in terms of the state labels
$\mu, \nu, \rho, \sigma$, the eigenvectors $C^{(sa)}_{M'}$ turn into
matrices and Eq.~(\ref{eq3.9}) contains a partial trace, 
\begin{equation}
\label{eq3.10}
\sum_{\rho \sigma} A^{(k)}_{\mu \nu, \rho \sigma}(2) C^{(sa)}_{\sigma
  \rho} = \Lambda^{s}(k) C^{(sa)}_{\mu \nu} \ .
\end{equation}
The eigenvectors can be orthonormalized (again this involves a trace),
\begin{equation}
\label{eq3.11}
\sum_{\mu \nu} C^{(sa)}_{\mu \nu} C^{(tb)}_{\nu \mu} = N \delta_{st}
\delta_{ab} \ . 
\end{equation}
For reasons which will be obvious later, we have chosen the
normalization constant as $N$ rather than unity. The eigenvalue
expansion of $A^{(k)}_{\mu \nu, \rho \sigma}(2)$ has the form
\begin{equation}
\label{eq3.12}
A^{(k)}_{\mu \nu, \rho \sigma}(2) = \frac{1}{N} \sum_{sa}
\Lambda^{(s)}(k) C^{(sa)}_{\mu \nu} C^{(sa)}_{\rho \sigma} \ . 
\end{equation}
We proceed to construct the eigenvalues $\Lambda^{(s)}(k)$ and
eigenvectors $C^{(sa)}_{\mu \nu}$.

In searching for the solutions of Eq.~(\ref{eq3.10}), we are guided by
two simple examples. The first one is the case of a random potential
$V({\vec x})$. Let us suppose that the potential has a Gaussian
distribution with zero mean value and a second moment given by
$\overline{V({\vec x}) V({\vec y})} = {\cal F}({\vec x},{\vec y})$. We
assume that translational and rotational invariance hold and that,
therefore, ${\cal F}$ is a function only of $|{\vec x} - {\vec   y}|$.
Then, it is possible and useful to decompose ${\cal F}$ into
invariants with respect to the rotational group. This is accomplished
with the help of the multipole expansion
\begin{equation}
\label{eq3.13}
{\cal F}(|{\vec x} - {\vec y}|) = \sum_{L,M} f_L(|{\vec x}|,|{\vec
  y}|) Y_L^M(\Theta_x, \Phi_x) (Y_L^M(\Theta_y,\Phi_y))^* \ . 
\end{equation}
We have used polar coordinates, and the $Y_{L,M}$'s are the standard
spherical harmonics. The term $\sum_M Y_L^M(\Theta_x, \Phi_x)
(Y_L^M(\Theta_y,\Phi_y))^*$ is the projector onto an invariant
subspace. The spherical harmonics are eigenvectors of ${\cal F}$ with
eigenvalue $f_L$ and span the invariant subspace. We look for an
analogous decomposition of $A_{M M'}^{(k)}$, using the fact that
EGUE($k$) is invariant under unitary transformations of the
single--particle basis. We expect that $M$ and $M'$ play the role of
the polar angles $(\Theta_x, \Phi_x)$ (or $(\Theta_y,\Phi_y)$,
respectively), and that the projector $\sum_M Y_L^M(\Theta_x, \Phi_x)
(Y_L^M(\Theta_y,\Phi_y))^*$ is replaced by a projector onto a subspace
of Hilbert space which is invariant under unitary transformations of
the single--particle states $1,\ldots,l$.

Our second example is the GUE. The second moment of the GUE
interaction has the form of Eq.~(\ref{eq3.4}) with $\beta = 2$. The
two Kronecker symbols $\delta_{\mu \nu} \delta_{\rho \sigma}$ display
the unitary invariance of the GUE in Hilbert space. From the point of
view of our first example, we look at Eq.~(\ref{eq3.4}) as kin to a
multipole expansion. The expansion has only a single term, and the
product of the two Kronecker symbols is the projector onto an
invariant subspace. The subspace is spanned by eigenfunction(s) of the
second moment. There is one eigenfunction which belongs to the
eigenvalue $\lambda^2 = N$. It is the matrix $\delta_{\sigma \rho}$. All
traceless matrices are likewise eigenfunctions but belong to
eigenvalue zero.

We turn to Eq.~(\ref{eq3.10}). To construct the eigenvectors, we state
a fact implied by our derivation of the duality relation: For
$A^{(k)}(2)$ not to vanish, the two states $|\sigma \rangle$ and $|
\rho \rangle$ may differ in the occupation of at most $(m-k)$
single--particle states, and likewise for $| \mu \rangle$ and $|\nu
\rangle$. Guided by this observation and by the example of the GUE, we
begin with the simplest case denoted by $s = 0$. It is easy to check
that $C^{(0)}_{\mu \nu} = \delta_{\mu \nu}$ is a normalized
eigenvector of $A^{(k)}(2)$ with eigenvalue $\Lambda^{(0)}(k) = {m
  \choose k}{l-m+k \choose k}$. For $s = 1$, it is likewise easy to
check that for $j \neq i$, the vectors $C^{(1a)}_{\mu \nu}$ given by
$\langle \mu | a_j^{\dagger} a_i | \nu \rangle$ obey
Eq.~(\ref{eq3.10}) with eigenvalue $\Lambda^{(1)}(k) = {m-1 \choose
  k}{l-m+k-1 \choose k}$. For $j = i$, an additional construction is
needed because $\langle \mu | \sum_{j=1}^l a_j^{\dagger} a_j | \nu
\rangle = m \delta_{\mu \nu}$ so that this particular linear
combination reduces to a multiple of the eigenvector $C^{(0)}_{\mu
  \nu}$. We must, in fact, construct traceless matrices. For $j =
1,\ldots,l$ these are given by $\langle \mu | a_j^{\dagger} a_j -
(1/l) \sum_p a_p^{\dagger} a_p | \nu \rangle$. They are eigenfunctions
of $A^{(k)}(2)$ with eigenvalue $\Lambda^{(1)}(k)$. Only $(l-1)$ of
them are linearly independent, however. We observe that the eigenvalue
$\Lambda^{(1)}(k)$  has degeneracy $D^{(1)} = (l^2 - 1)$. The
operators $a_j^{\dagger} a_i$ defining the eigenvectors $C^{(1a)}_{\mu
  \nu}$ span a linear space which is invariant under unitary
transformations of the single--particle basis. (Trivially
$\sum_{j=1}^l a_j^{\dagger} a_j$ is also an invariant).

It remains to find linear combinations of the eigenvectors
$C^{(1a)}_{\mu \nu}$ which obey the orthonormality
conditions~(\ref{eq3.11}). For $j \neq i$, we consider the matrices
$\langle \mu | (a_j^{\dagger} a_i + a_i^{\dagger} a_j) | \nu \rangle$
and $\langle \mu | \sqrt{-1}(a_j^{\dagger} a_i - a_i^{\dagger} a_j) |
\nu \rangle$ of two Hermitean operators. These matrices are all
orthogonal upon each other. Normalization is assured by multiplying
each matrix with $\sqrt{(1/2)l(l-1)/(m(l-m))}$. For $j = i$, the
matrices have the form $\langle \mu | a_j^{\dagger} a_j - (1/l) \sum_p
a_p^{\dagger} a_p | \nu \rangle = \langle \mu | B_j | \nu \rangle$.
These matrices are automatically orthogonal to the ones with $j \neq
i$. We note that for $j,i = 1,\ldots,l$, the matrix $T_{ij} = (1/N)
{\rm tr} ( B_i B_j )$ has the form $(m/l) ((l-m)/(l-1)) [{\bf 1}
- (1/l) {\bf M}]$. Here, ${\bf 1}$ is the unit matrix in $l$
dimensions, and the $l$--dimensional matrix ${\bf M}$ carries the
entry $(+1)$ in every element. The matrix $T_{ij}$ has the
non--degenerate eigenvalue zero (which arises because $\sum_j B_j =
0$) and the $(l-1)$--fold degenerate eigenvalue $(m/l) (l-m)/(l-1)$.
Transforming $B_j$ with the diagonalizing orthogonal matrix and
multiplying the result with $\sqrt{[l (l-1)] / [m (l-m)]}$ yields
$(l-1)$ orthonormal eigenvectors. Thus, we have explicitly constructed
the $D^{(1)}$ orthonormal eigenvectors to eigenvalue $\Lambda^{(1)}(k)$.
We observe that these do not depend on the rank $k$ of the interaction.
The rank affects only the value of $\Lambda^{(1)}(k)$. In order to
simplify our notation, we will continue to denote these orthonormal
eigenvectors by $C^{(1a)}_{\mu \nu}$ where $a = 1,\ldots,D^{(1)}$.

It should now be clear how to proceed: The index $s$ labels the rank
of the operators appearing in the matrices $C^{(sa)}_{\mu \nu}$. It is
easy to check that for \{$j_1,\ldots,j_s$\} with $j_1 < \ldots < j_s$
different from \{$i_1,\ldots,j_s$\} with $i_1 < \ldots < i_s$ (so that
no two indices are equal), the matrices $\langle \mu |
a_{j_1}^{\dagger} \ldots a_{j_s}^{\dagger} a_{i_s} \ldots a_{i_1} |
\nu \rangle$ are eigenvectors of $A^{(k)}(2)$ with eigenvalue
\begin{equation}
\label{eq3.14}
\Lambda^{(s)}(k) = {m-s \choose k} {l-m+k-s \choose k} \ .
\end{equation}
These matrices are orthogonal upon all matrices in classes $s' \neq
s$. They can be made mutually orthogonal by using the Hermitean
combinations of operators $( a_{j_1}^{\dagger} \ldots
a_{j_s}^{\dagger} a_{i_s} \ldots a_{i_1} + a_{i_1}^{\dagger} \ldots
a_{i_s}^{\dagger} a_{j_s} \ldots a_{j_1} )$ and $\sqrt{-1}
(a_{j_1}^{\dagger} \ldots a_{j_s}^{\dagger} a_{i_s} \ldots a_{i_1} -
a_{i_1}^{\dagger} \ldots a_{i_s}^{\dagger} a_{j_s} \ldots a_{j_1} )$.
This is the same construction as used for $s = 1$. Whenever at
least two indices from the sets \{$j_1 < \ldots < j_s$\} and \{$i_1 <
\ldots < j_s$\} are equal, the same problem as for $s=1$ arises.
Indeed, given a matrix in class $s' < s$, we can seemingly ``lift'' it
into class $s$ by multiplying the defining operator by $[ \sum_{j=1}^l
a_j^{\dagger} a_j ]^{(s - s')}$, even though the operator
$\sum_{j=1}^l a_j^{\dagger} a_j$ can be replaced by $m$. Therefore, it
is necessary to find linear combinations which are orthogonal upon
these constructs. It is not necessary to perform the construction
explicitly. Suffice it to say that the dimension $D^{(s)}$ of the 
resulting linear space is given by $D^{(0)} = 1$ and, for $s \ge 1$ by
\begin{equation}
\label{eq3.15}
D^{(s)} = {l \choose s}^2 - {l \choose s-1}^2 \ .
\end{equation}
This is because the total number of sets \{$j_1 < \ldots < j_s$\} and
\{$i_1 < \ldots < j_s$\} is ${l \choose s}^2$, and because we have to
subtract the number of matrices obtained by ``lifting'' from all
lower classes. We denote the orthonormal eigenvectors in class $s$ by
$C^{(sa)}_{\mu \nu}$ where $a = 1,\ldots,D^{(s)}$. Again, the
eigenvectors do not depend upon the rank $k$ of the interaction, only
the eigenvalue does.

The total number of eigenvectors constructed in this fashion is given
by $\sum_{s=0}^{m} D^{(s)} = {l \choose m}^2$. This is the dimension
of the grand matrix $A^{(k)}(2)$. We conclude that we have found a
complete orthonormal set of eigenvectors obeying Eq.~(\ref{eq3.11}),
and that, therefore, the expansion~(\ref{eq3.12}) applies. We recall
that the rank $k$ of the interaction appears only in the eigenvalues
given by Eq.~(\ref{eq3.14}). Inspection of this equation shows that
the sum in Eq.~(\ref{eq3.12}) actually terminates at $s = (m-k)$. For
$k = m$ we retrieve the GUE where only $s = 0$ occurs, and where
$\Lambda^{(0)}(m) = {l \choose m} = N$, in keeping with
Eq.~(\ref{eq3.4}).  

For the orthogonal case, we observe that the matrix $A^{(k)}(1)$ given
by Eq.~(\ref{eq3.3}) is the sum of two terms each of which have the
form of $A^{(k)}(2)$. We apply the arguments for $A^{(k)}(2)$ to each
of these terms separately and find as a result
\begin{equation}
\label{eq3.16}
A^{(k)}_{\mu \nu, \rho \sigma}(1) = \frac{1}{N} \sum_{sa}
\Lambda^{(s)}(k) [ C^{(sa)}_{\mu \nu} C^{(sa)}_{\rho \sigma} +
C^{(sa)}_{\mu \rho} C^{(sa)}_{\nu \sigma} ] \ .
\end{equation}
The eigenvalues and eigenvectors are the same as for the case $\beta =
2$. We can combine Eqs.~(\ref{eq3.12}) and (\ref{eq3.16}) into the
single equation
\begin{equation}
\label{eq3.17}
A^{(k)}_{\mu \nu, \rho \sigma}(\beta) = \frac{1}{N} \sum_{sa}
\Lambda^{(s)}(k) [ C^{(sa)}_{\mu \nu} C^{(sa)}_{\rho \sigma} +
\delta_{\beta 1} C^{(sa)}_{\mu \rho} C^{(sa)}_{\nu \sigma} ] \ .
\end{equation}
Eqs.~(\ref{eq3.11}), (\ref{eq3.14}), (\ref{eq3.15}) and (\ref{eq3.17}),
constitute the central results of this Section.

Another important identity is obtained by combining the eigenvalue
expansion for $A^{(k)}(2)$, Eq.~(\ref{eq3.10}), with the duality
relation~(\ref{eq3.5}). This yields
\begin{equation}
\label{eq3.18}
\sum_{sa} \Lambda^{(s)}(k) C^{(sa)}_{\mu \nu} C^{(sa)}_{\rho \sigma} =
\sum_{tb} \Lambda^{(t)}(m-k) C^{(tb)}_{\mu \sigma} C^{(tb)}_{\rho \nu}
\ .
\end{equation}
This equation implies further useful relations. For instance, putting
$\rho = \nu$ and $\mu = \sigma$, summing over $\mu$ and $\rho$, and
using the orthonormality relations we find
\begin{equation}
\label{eq3.19}
\sum_s \Lambda^{(s)}(k) D^{(s)} = N \Lambda^{(0)}(m-k) \ .
\end{equation}

\section{Moments of $V_k$}
\label{low}

We apply the results obtained in Sections~\ref{bas} and \ref{eig} and
calculate the dependence on $k,m$ and $l$ of measures of the shape of
the average spectrum of EGOE($k$) and EGUE($k$). As always in this
paper, we do so in the limit $l \rightarrow \infty$. Only in this
limit do we expect to obtain generic results. This limit corresponds
to the limit $N \rightarrow \infty$ of canonical random--matrix theory.
As a by--product, we obtain results which allow us to infer how
quickly the embedded ensembles approach the limit $l \rightarrow
\infty$.

We calculate ensemble--averages over products of traces of powers of
$V_k(\beta)$. Traces are always taken in the Hilbert space of Slater
determinants $| \mu \rangle$. From Eqs.~(\ref{eq3.17})
and (\ref{eq3.19}) we have
\begin{equation}
\label{eq4.1}
\overline{(\frac{1}{N} {\rm tr} V_k(\beta))^2} = \frac{(1 +
  \delta_{\beta 1})}{N^2} \sum_{s=0}^{m-k} \Lambda^{(s)}(k) D^{(s)} =
  \frac{(1 + \delta_{\beta1})}{N} \Lambda^{(0)}(m-k)
\end{equation}
and
\begin{equation}
\label{eq4.2}
\frac{1}{N} {\rm tr} \overline{(V_k(\beta))^2} = \Lambda^{(0)}(k) +
\delta_{\beta 1} {\cal O}(1/l^2) \ .
\end{equation}
Two further averages are given by
\begin{eqnarray}
\label{eq4.3}
\overline{\biggl[\frac{1}{N} {\rm tr} (V_k(\beta))^2\biggr]^2} &=&
(\Lambda^{(0)}(k))^2 + \frac{2 (1 + \delta_{\beta 1})}{N^2}
\sum_{s=0}^{m-k} (\Lambda^{(s)}(k))^2 D^{(s)} 
\nonumber \\
&&\qquad + \delta_{\beta 1} {\cal O}(1/l^2) 
\nonumber \\
&=& (\Lambda^{(0)}(k))^2 + \frac{2 (1 + \delta_{\beta 1})}{N^2}
\sum_{s=0}^{k} (\Lambda^{(s)}(m-k))^2 D^{(s)} 
\nonumber \\
&&\qquad + \delta_{\beta 1} {\cal O}(1/l^2)
\end{eqnarray}
and
\begin{eqnarray}
\label{eq4.4}
\frac{1}{N} {\rm tr} \overline{(V_k(\beta))^4} &=& 2
(\Lambda^{(0)}(k))^2 + \frac{1}{N} \sum_{s=0}^{{\rm min}(k,m-k)}
\Lambda^{(s)}(k) \Lambda^{(s)}(m-k) D^{(s)} \nonumber \\
&& \qquad + \delta_{\beta 1} {\cal O}(1/l^2) \ .
\end{eqnarray}
Both these results can easily be found for $\beta = 2$. For $\beta =
1$, we encounter terms of the form $\sum_{\nu} C^{(sa)}_{\mu \nu} 
C^{(sa)}_{\rho \nu}$ which cannot be worked out with the help of the
orthonormality relation~(\ref{eq3.11}). We recall the explicit
construction of the matrices $C^{(sa)}_{\mu \nu}$ in Section~\ref{eig}.
When the indices $j_1,\ldots,j_s$ and $i_1,\ldots,i_s$ of the operator
defining the matrix $C^{(sa)}_{\mu \nu}$ are all different, the
Hermitean operators come in pairs of the form $(A + A^{\dagger})$ and
$i(A - A^{\dagger})$ where $A = (a_{j_1}^{\dagger} \ldots
a_{j_s}^{\dagger} a_{i_s} \ldots a_{i_1} )$. The matrix $C^{(sa)}_{\mu
  \nu}$ containing $(A + A^{\dagger})$ is real and symmetric, the one
containing $i(A - A^{\dagger})$ is purely imaginary and antisymmetric.
Using these symmetry properties and completeness, we find that in the
sum $\sum_{\nu} C^{(sa)}_{\mu \nu} C^{(sa)}_{\rho \nu}$, the two
operators $A$ and $A^{\dagger}$ appear only in the forms $(A)^2$ and
$(A^{\dagger})^2$. These terms both vanish because $A$ is
nilpotent. We are left with the contributions where at least one pair
of indices $j_1,\ldots,j_s$ and $i_1,\ldots,i_s$ of the operator
defining the matrix $C^{(sa)}_{\mu \nu}$ coincides. The number of such
terms in the sum over $a$ is of relative order $1/l^2$ and, thus,
negligible in comparison with other terms which we keep.

We construct three ratios which yield information about the shape of
the spectrum. The ratio $S$ measures the fluctuations of the center of
the spectrum in units of the average width of the spectrum.
It is given by
\begin{equation}
\label{eq4.5}
S = \frac{\overline{\biggl(\frac{1}{N} {\rm tr} V_k(\beta)
    \biggr)^2}}{\frac{1}{N} {\rm tr} \overline{(V_k(\beta))^2}} \ .
\end{equation}
The relative fluctuation of the width of the spectrum is given by
\begin{equation}
\label{eq4.6}
R = \frac{\overline{ \biggl( \frac{1}{N} {\rm tr} (V_k(\beta))^2
  \biggr)^2} - \biggl(\frac{1}{N} {\rm tr} \overline{
  (V_k(\beta))^2}\biggr)^2} {\biggl( \frac{1}{N} {\rm tr} \overline{
  (V_k(\beta))^2}\biggr)^2} \ .
\end{equation}
The kurtosis $\kappa$ is given by
\begin{equation}
\label{eq4.7}
\kappa = \frac{\frac{1}{N} {\rm tr} \overline{ (V_k(\beta))^4}}
  {\biggl( \frac{1}{N} {\rm tr} \overline{ (V_k(\beta))^2}\biggr)^2} =
  2 + Q \ .
\end{equation}
The quantity $Q$ marks the difference in spectral shape between the
semicircle $(\kappa = 2)$ and the Gaussian ($\kappa = 3$).

The ratios $S$ and $R$ yield information about finite--size effects in
numerical calculations of EGOE($k$) and EGUE($k$)~\cite{flo00}. 
%%%
The ratio $S$ is given by $S(k,m,l) = ((1 + \delta_{\beta 1})/N)
(\Lambda^{(0)}(m-k) / \Lambda^{(0)}(k))$. The ratio $R$ has the value 
\begin{eqnarray}
\label{eq4.8}
R(k,m,l) &=& \frac{2(1 + \delta_{\beta 1})} {N^2} \sum_{s=0}^{m-k}
\biggl(\frac{\Lambda^{(s)}(k)} {\Lambda^{(0)}(k)}\biggr)^2 D^{(s)}
\nonumber \\ 
&=&\frac{2(1 + \delta_{\beta 1})}{N^2} \sum_{s=0}^{k} \biggl(
\frac{\Lambda^{(s)}(m-k)} {\Lambda^{(0)}(k)}\biggr)^2 D^{(s)} \ .
\end{eqnarray}
The result for $R$ is consistent with the estimate by
French~\cite{fre73}.  For $k \geq 1$, we have both $S \rightarrow 0$
and $R \rightarrow 0$ for $l \rightarrow \infty$.  However, for fixed
values of $k$ and $f$, $S \propto l^{-2k} \propto (\ln N)^{-2k}$ and
$R \propto l^{-2k} \propto (\ln N)^{-2k}$ decrease very slowly as $l$
increases, much more slowly than in the case of the canonical
ensembles where $S(m,m,l) = 1/N^2$ and $R(m,m,l) = 2/N^2$.
%%%
This fact illustrates the difficulty in obtaining reliable spectral
information numerically. We note that $R(0,m,l) = 2$ and return to
this point in Section~\ref{bin} in the context of the binary
correlation approximation.

\begin{figure}
\begin{minipage}{15cm}
\centerline{\psfig{figure=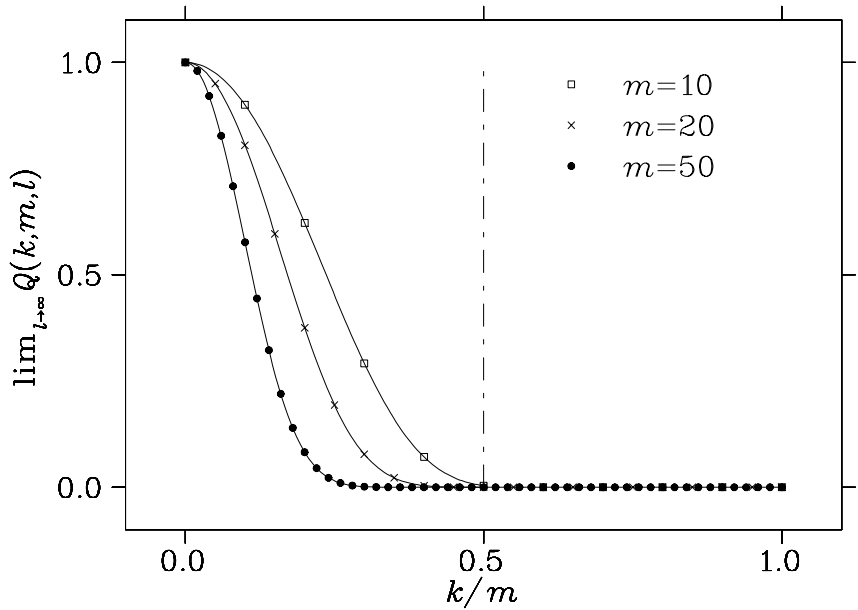,width=8cm,angle=90}}
\end{minipage}
\caption{The limit $l \rightarrow \infty$ of $Q(k,m,l)$ versus $k/m$
  for three values of $m$.}
\label{fig2}
\end{figure}

From the point of view of the present study, the most interesting
information resides in the ratio $Q$ defined in Eq.~(\ref{eq4.7}) and
given by
\begin{equation}
\label{eq4.9}
Q(k,m,l) = \frac{1}{N} \sum_{s=0}^{{\rm min}(m-k,k)} 
\frac{\Lambda^{(s)}(k) \Lambda^{(s)}(m-k)} {(\Lambda^{(0)}(k))^2}
D^{(s)} \ .
\end{equation}
We have $Q(0,m,l) = 1$ and $Q(m,m,l) = 1/N^2$, consistent with a
Gaussian and, for $N \rightarrow \infty$, a semicircular shape of the
spectrum, respectively. Moreover, recalling that $f = m/l$, we have
the following limits:
\begin{eqnarray}
\label{eq4.10}
{\rm lim}_{l \rightarrow \infty} Q &=& 0 \ , \qquad \ \ {\rm for} \ 
  2k > m \ {\rm with \ both} \ k \ {\rm and} \ m \ {\rm fixed} \ ,
\nonumber \\
{\rm lim}_{l \rightarrow \infty} Q &=& \frac{{m-k \choose k}}{{m
  \choose k}} \ , \ {\rm for} \ 2k \leq m \ {\rm with \ both} \ k \ {\rm
  and} \ m \ {\rm fixed} \ ,
\nonumber \\ 
{\rm lim}_{l \rightarrow \infty} Q &=& 0 \ , \ \ {\rm with \ both } \
  k/m \ {\rm and} \ f \ {\rm fixed} \ ,
\nonumber \\ 
{\rm lim}_{l \rightarrow \infty} Q &=& 1 \ , \ \ {\rm with \ both } \ k
  \ {\rm and} \ f \ {\rm fixed} \ .
\end{eqnarray}
A graphical representation of ${\rm lim}_{l \rightarrow \infty} Q$
is given in Figure~\ref{fig2}. We note that as $m$ increases, the
curves shrink toward the ordinate. In the limit $m \rightarrow
\infty$, the value of $Q$ is unity at $k/m = 0$ and zero everywhere
else.

These results show that in the limit $l \rightarrow \infty$ and for
fixed $m$, the spectral shape is that of a semicircle as long as $2k >
m$. (In the unphysical limit where with $l$ also $k$ tends to
infinity, we also obtain a semicircle). The semicircle undergoes a
smooth transition to Gaussian shape for $2k \leq m$. We have
identified the critical value of $k$ where the transition of the
average spectral shape from semicircle to Gaussian begins. Obviously,
the kurtosis alone is not sufficient proof for this transition: We
have not investigated the higher moments of the spectrum. The
calculation of such moments is very difficult at best. In
Section~\ref{sup} we present independent evidence for the conclusion
that the transition to Gaussian shape begins at $2k = m$.

For $k = m$, we have $Q \propto 1/N$ for $l \rightarrow \infty$. For
small $k$, the approach is not quite as rapid. For instance, for $k =
1$, we have $Q \propto l^3 / [ N m (l-m) ]$.

\section{Supersymmetry Approach}
\label{sup}

Until now, the embedded ensembles have largely resisted attempts to
apply the supersymmetry method~\cite{ver84}. This is because these
ensembles lack the orthogonal (unitary) invariance in Hilbert space
characteristic of GOE (GUE, respectively). The eigenvector
expansion~(\ref{eq3.12}) remedies this deficit: After averaging over
the ensemble, the integrand of the generating functional contains an
exponential the argument of which is a sum of squares of bilinear
forms in the integration variables. This expression can be linearized
via the familiar Hubbard--Stratonovich transformation. We use the
saddle--point approximation, solve for the saddle--point manifold, and
consider the lowest--order terms in the loop expansion for the
one--point and the two--point functions. The results in
Sections~\ref{loop1} and \ref{loop2} supersede and correct statements
made in Section 5 of Ref.~\cite{wei00}. In a last Section, we show
that the spectral fluctuations of the limiting ensembles EGOE$_{\rm
  max}(k$) and EGUE$_{\rm max}(k$) defined in Section~\ref{lim}
coincide with those of the GOE and GUE, respectively. For simplicity,
we display only the case $\beta = 2$. The calculations for $\beta = 1$
can be carried through in full analogy.

\subsection{Saddle Point}
\label{sad}

The steps leading to the saddle--point equation are perfectly
standard~\cite{efe83,ver85} and are only sketched here. We consider
the case of the two--point function and obtain the one--point function
later by specialization. The integrand of the generating functional
contains the factor
\begin{equation}
\label{eq6.1}
\exp \ [ \ \frac{i}{2} \sum_{\mu \nu p \alpha} \Psi_{\mu p \alpha}^*
L_{p \alpha}^{1/2} \langle \mu | V_k | \nu \rangle L_{p \alpha}^{1/2}
\Psi_{\nu p \alpha} \ ] \ .
\end{equation}
Here, $\Psi_{\mu p \alpha}$ stands for the complex integration
variables with $\alpha = 0,1$ for commuting (anticommuting) variables,
$p = 1,2$ for the advanced (retarded) Green function, and $\mu =
1,\ldots,N$. The diagonal supermatrix $L$ distinguishes the retarded
and advanced case and is defined in Ref.~\cite{ver85}. Averaging
expression~(\ref{eq6.1}) over the ensemble and using Eq.~(\ref{eq3.1}),
we obtain a quartic term in the $\Psi$'s involving the matrix
$A^{(k)}$. We use the eigenvector decomposition~(\ref{eq3.12}). This
allows us to perform the Hubbard--Stratonovich transformation. For
each value of the pair $(s,a)$, we introduce a supermatrix $\sigma^{(s
  a)}$ of composite variables. Integration over the $\Psi$'s yields an
integral over the composite variables containing the factor
$\exp(-{\cal L}_{\rm eff})$. The effective Lagrangean ${\cal L}_{\rm
  eff}$ has the form
\begin{equation}
\label{eq6.2}
\frac{N}{2} \sum_{s a} {\rm trg} [ \sigma^{(s a )} ]^2 +
  {\rm tr}_{\mu} {\rm trg} \ln \biggl[(E - \frac{1}{2}{\epsilon} L)
  \delta_{\mu \nu} - \sum_{s a} \lambda^{(s)}(k) \sigma^{(s a)}
  C^{(s a)}_{\mu \nu} - J_{\mu \nu} \bigg] \ .
\end{equation}
The $\lambda^{(s)}(k)$'s are the positive square roots of the
eigenvalues $\Lambda^{(s)}(k)$. The energy arguments $E_1$ and $E_2$
of the advanced and the retarded Green function are used to define $E
= (1/2)(E_1 + E_2)$ and $\epsilon = E_2 - E_1$ while $J$ stands for
the source terms. The occurrence of the factor $N \gg 1$ in the
effective Lagrangean causes us to use the saddle--point approximation.
The wish to make this factor explicit was the reason for the
normalization chosen in Eq.~(\ref{eq3.12}). Under omission of the
small terms $J$ and ${\epsilon \propto 1/N}$, the saddle--point
equations read
\begin{equation}
\label{eq6.3}
\sigma^{(s a)} = \frac{1}{N} {\rm tr}_{\mu} \biggl( [ E - \sum_{t
  b} \lambda^{(t)}(k) \sigma^{(t b)} C^{(t b)} ]^{-1} \lambda^{(s)}(k)
  C^{(s a)} \biggr) \ .
\end{equation}
We have used matrix notation in Hilbert space. The saddle--point
Eqs.~(\ref{eq6.3}) constitute a set of coupled equations, one for each
of the unknown quantities $\sigma^{(s a)}$.

To solve the saddle--point equations~(\ref{eq6.3}), we multiply the
equation for $\sigma^{(s a)}$ with $\lambda^{(s)}(k) \ C^{(s a)}$ and
sum over all $(s, a)$. We find 
\begin{equation}
\label{eq6.4}
X_{\rho \sigma} = \frac{1}{N} {\rm tr}_{\mu} \biggl[ (E - X)^{-1}
\sum_{s a} \Lambda^{(s)}(k) C^{(s a)} \biggr] C^{(s a)}_{\rho \sigma}
\end{equation}
where $X_{\rho \sigma} = \sum_{s a} \lambda^{(s)}(k) \sigma^{(s a)}
C^{(s a)}_{\rho \sigma}$. Comparison with Eq.~(\ref{eq3.12}) shows
that the right--hand side of Eq.~(\ref{eq6.4}) can be expressed in
terms of the grand matrix $A^{(k)}$. Doing so and using
Eq.~(\ref{eq3.1}), we obtain 
\begin{equation}
\label{eq6.5}
X_{\rho \sigma} = \sum_{\mu \nu} \overline{ \langle \rho | V_k | \mu
  \rangle [ (E - X)^{-1}]_{\mu \nu} \langle \nu | V_k | \sigma
  \rangle} \ .
\end{equation}
The overline refers exclusively to the two matrix elements of $V_k$
displayed explicitly. We use Eq.~(\ref{eq6.5}) in the identity
$\overline{G(E)} = [E - X]^{-1}$ for the averaged Green function (a
one--point function). This yields the generalization of the Pastur
equation to EGUE($k$). Iteration with respect to $X$ of this equation
yields a series which can be interpreted in terms of diagrams: Every
averaged pair of $V_k$'s is connected by a contraction line. The
series differs from the series obtained by averaging the Born series
for $G(E)$. It lacks all those contributions where at least two
contraction lines intersect. We return to this point in
Section~\ref{con}. Terminating the iteration after a finite number of
steps (this number is arbitrary), we find that $X$ obeys the GUE
saddle--point equation $X = \Lambda^{(0)}(k) [ E - X ]^{-1}$. We first
look for a solution $X^{\rm diag}$ which is diagonal in the
superindices. For $|E| \leq 2 \lambda^{(0)}(k)$ this yields $X_{\mu
  \nu}^{\rm diag} = \delta_{\mu \nu} \lambda^{(0)}(k) \tau^{(0)}$ with
\begin{equation}
\label{eq6.6}
\tau^{(0)} = \frac{E}{2 \lambda^{(0)}(k)} \pm i \sqrt{1 -
  \biggl(\frac{E}{2 \lambda^{(0)}(k)}\biggr)^2} \ .
\end{equation}
The $\pm$ signs refer to the retarded ($G^{-}(E)$) and the advanced
($G^{+}(E)$) case, respectively. Inserting this result back into the
saddle--point equations~(\ref{eq6.3}), we find that $\sigma^{(0)}$ is
identical to $\tau^{(0)}$ while $\sigma^{(s a)} = 0$ for all $s \geq
1$. For the one--point function this shows that, within the range of
validity of the saddle--point approximation, the semicircle describes
the generic form of the average spectrum of EGUE($k$). In contrast to
the standard GUE case, the radius $\lambda^{(0)}(k)$ of the semicircle
tends to infinity for $l \rightarrow \infty$. The mean level spacing
$d(k) \propto \lambda^{(0)}(k) / N$ tends to zero ($m$ fixed) or
remains finite ($f$ fixed) in the same limit.

In the case of the canonical ensembles, the invariance of the
effective Lagrangean under general pseudounitary transformations
implies that the two--point function possesses not a single
saddle point but a saddle--point manifold. The same situation
prevails here. The manifold is given by $T^{-1} \sigma^{(0)} T$. Here
$\sigma^{(0)}$ is a diagonal supermatrix of dimension four with
entries given by Eq.~(\ref{eq6.6}) in the usual way. The matrices $T$
parametrize the coset space UOSP$(1,1/1,1) / [$UOSP$(1,1) \otimes
$UOSP$(1,1)]$. Using this fact and the saddle--point values
$\sigma^{(sa)} = 0$ for $s \geq 1$ in the effective Lagrangean, we
find that the first--order term in $\epsilon$ takes the canonical form
\begin{equation}
\label{eq6.6a}
- \frac{i \pi \epsilon}{d} {\rm trg} ( L T^{-1} L T ) \ .
\end{equation}
Here $d$ is the average level spacing. The argument also carries
through for higher--point correlation functions and shows that the
spectral fluctuations of EGUE($k$) are identical to those of the GUE.

\subsection{Loop Correction to the One--Point Function}
\label{loop1}

We wish to determine the range of validity of the saddle--point
solution in the limit $l \rightarrow \infty$. We do so by using a
power--series expansion for the effective Lagrangean around the
saddle--point and a subsequent expansion of the resulting exponential
(``loop expansion''). We recall that in the GUE case, all terms of the
resulting series save the first vanish individually as the dimension
$N$ of the GUE matrix tends to infinity. This is true except at the
edge of the spectrum. In the present case, the calculation of the
terms in the loop expansion is a formidable task, and we confine
ourselves to the correction term of lowest non--vanishing order. We
investigate the behavior of this term for $l \rightarrow \infty$.

For the one--point function, the sign in Eq.~(\ref{eq6.6}) is fixed
(negative, say), the dimension of the supermatrices is two, and we deal
with a single energy variable $E_1$ only. We write $\sigma^{(sa)} =
\delta_{s 0} \tau^{(0)} + \delta \sigma^{(sa)}$ and expand the
effective Lagrangean in powers of $\delta \sigma^{(sa)}$. For the
logarithmic term, we effectively generate an expansion in powers of
$\delta \Sigma = \sum_{s a} \lambda^{(s)}(k) C^{(sa)} \delta
\sigma^{(sa)}$. In this expansion, we keep terms up to fourth order.
We expand each of the resulting expressions up to first order in the
source variable $J$, and the exponential in powers of the terms so
generated. Details are given in Appendix~\ref{loop}. We obtain
%%%%%%%%%%%%%%%%%%%%%%%%%%%%%%%%%%%%%%%%%%%%%%%%%%%%%%%%%%%%%%%%%%%%%%
\begin{equation}
\label{eq6.7}
\overline{G^{+}(E)} = \frac{N \tau^{(0)} {\rm trg} J}{
  \lambda^{(0)}(k)} \left( 1 + \frac{(\tau^{(0)})^4}{N(\tilde{Y}(k))^2} 
  \sum_{s} \frac{\Lambda^{(s)}(k)
  \Lambda^{(s)}(m-k)}{(\Lambda^{(0)}(k))^2} D^{(s)} \right) \ .
\end{equation}
The squared factor $\tilde{Y}(k)$ is some suitably chosen complex
number depending on $k$, $m$, $l$, and $E$. $(\tilde{Y}(k))^{-2}$
compensates for dropping factors $Y^{(s)}(k) = 1 - (\tau^{(0)})^2
\Lambda^{(s)}(k) / \Lambda^{(0)}(k)$ in the denominators inside the
sum in Eq.~(\ref{eq6.7}). These factors $Y^{(s)}(k)$ originate from
the Gaussian weight factors. $\tilde{Y}(k)$ is well defined with
$|\tilde{Y}(k)|$ finite as long as $|Y^{(s)}(k)|$ is finite for all
$s$. At the center of the spectrum where $(\tau^{(0)})^2 = (-1)$, we
have $1 < Y^{(s)}(k) < 2$. (More generally, we have $0 \leq
|Y^{(s)}(k)| \leq 2$ everywhere in the spectrum.) In this case,
$\tilde{Y}(k)$ is real and lies within the same bounds. For the loop
correction, this gives the upper bound $Q(k,m,l)$, and
Eq.~(\ref{eq6.7}) simplifies to
%%%%%%%%%%%%%%%%%%%%%%%%%%%%%%%%%%%%%%%%%%%%%%%%%%%%%%%%%%%%%%%%%%%%%%
\begin{equation}
\label{eq6.8}
|\overline{G^{+}(0)}| \leq \frac{N |{\rm trg} J|}{\lambda^{(0)}(k)} 
  \biggl( 1 + Q(k,m,l)
\biggr) \ . 
\end{equation}
The loop correction is bounded by the ratio $Q$ defined in
Eq.~(\ref{eq4.9}), cf. also the relations~(\ref{eq4.10}). This fact is
highly satisfactory: It agrees with our evaluation of the kurtosis in
Section~\ref{low} in the sense that non--vanishing corrections to the
semicircle shape occur for $l \rightarrow \infty$ whenever $2k \leq m$
while the semicircle form prevails for $2k > m$. In view of the
bounds $1 < |Y^{(s)}(k)| < 2$ on $Y^{(s)}(k)$, the asymptotic behavior
implied by the relations~(\ref{eq4.10}) holds true independently of the
estimate leading to the relation~(\ref{eq6.8}). And this behavior
prevails within the major part of the spectrum. With $z^{(s)}(k) =
\Lambda^{(s)}(k) / \Lambda^{(0)}(k)$ and $x = E/(2 \lambda^{(0)}(k))$,
this statement follows from $|Y^{(s)}(k)| = \sqrt{1 + 2 z^{(s)} ( 1 - 2
  x^2 ) + (z^{(s)})^2}$. The modification of the shape of the spectrum
due to the loop correction is determined by the energy dependence of
$\tau^{(0)}$ which contains the term $\sqrt{1 - x^2}$. 
%%%%%%%%%%%%%%%%%%%%%%%%%%%%%%%%%%%%%%%%%%%%%%%%%%%%%%%%%%%%%%%%%%%%%%
At the edge of the spectrum, $Y^{(0)}(k)$ approaches zero. So does
$\tilde{Y}(k)$ and a more detailed analysis would be required to see
whether the square--root singularity is asymptotically removed by the
loop correction. We have not pursued this point.
%%%%%%%%%%%%%%%%%%%%%%%%%%%%%%%%%%%%%%%%%%%%%%%%%%%%%%%%%%%%%%%%%%%%%%

We conclude that the average spectrum of EGUE($k$) has semicircle
shape for $2k > m$ while corrections to this form arise for $2k \leq
m$. These corrections may be substantial. Thus, the conclusion reached
in Section~\ref{low} gains independent and strong support by the
present analysis.

\subsection{Loop Correction to the Two--Point Function}
\label{loop2}

For the two--point function, we again expand around the saddle
point. The logarithmic term in the effective Lagrangean has the form
\begin{equation}
\label{eq6.9}
{\rm tr}_{\mu} {\rm trg} \ln \biggl[E - \lambda^{(0)} \sigma^{(0)} -
  \frac{1}{2}{\epsilon} T L T^{-1} - T J T^{-1} - \sum_{s a} \delta
  \Sigma^{(sa)} \bigg] \ .
\end{equation}
We have absorbed the matrices $T$ and $T^{-1}$ in the definition of
the four--by--four matrices $\delta \Sigma^{(sa)}$. Except for $s = 0$
these matrices have 16 independent variables. For $s = 0$, we have
$\delta \sigma^{(0)} = T \delta P T^{-1}$ where $\delta P$ commutes
with $L$. This is because the remaining degrees of freedom in $\delta
\sigma^{(0)}$ generate the saddle--point manifold.

We expand the logarithm in powers of $\delta \Sigma$ and keep terms up
to order four. We expand the resulting terms in powers of $1/2
\ \epsilon \ T L T^{-1} + T J T^{-1}$ and keep terms up to order two.
We expand the exponential in powers of these terms. After integrating
over the $\delta \sigma$'s and dropping terms that vanish as $1/N$ for
$N \rightarrow \infty$, we write the result as an exponential, see
Appendix~\ref{loop}. This exponential contains two terms. The first contains
the expression
\begin{equation}
\label{eq6.10}
- \frac{i \pi \epsilon} {d_{\rm l}} {\rm trg} ( L T^{-1} L T ) \ .
\end{equation}
Here $d_{\rm l}$ is the modified average level spacing which results
when the last term in Eq.~(\ref{eq6.7}) is included in the average
Green function. The index ${\rm l}$ stands for loop. This modification
is no more than a consistency check. Indeed, the graded trace in
Eq.~(\ref{eq6.10}) yields Wigner--Dyson spectral fluctuations. These,
however, must be expressed in terms of the correct average level
spacing which, in the framework of the loop expansion, is precisely
$d_{\rm l}$.

Interest focusses on the second term. It is proportional to
%%%
\begin{equation}
\label{eq6.11}
R(k,m,l) \ \biggl( \frac{1}{d_{\rm l}} \ {\rm trg} \ \bigl[ \frac{1}{2}
\epsilon L T L T^{-1} + L T J T^{-1} \bigr] \biggr)^2 \ .   
\end{equation}
%%%
In a different context it was shown by Kravtsov and Mirlin~\cite{kra94}
that such a term produces deviations from Wigner--Dyson spectral
fluctuations. The strength of these fluctuations is proportional to
the ratio $R(k,m,l)$ given in Eqs.~(\ref{eq4.8}). We recall that $R$
vanishes for $l \rightarrow \infty$ but does so particularly slowly
for small values of $k$ and fixed $f$.

Whereas the lowest non--vanishing loop correction has given us clear
evidence for a change in spectral shape at $2k = m$, this is not the
case for the spectral fluctuation properties. The ratio $R$ vanishes
asymptotically more slowly than $1/N$ but does not indicate a break at
$2k = m$. If the transition to Gaussian shape is accompanied by a
change in level statistics, we would expect that terms like the one in
(\ref{eq6.11}) but which do not vanish asymptotically, would arise in
higher--order terms of the loop expansion. We return to this point in
Section~\ref{bin1}.

\subsection{Universality of EGUE$_{\rm max}(k)$}
\label{max}

To show universal random--matrix behavior for EGOE$_{\rm max}(k)$, we
apply the supersymmetry formalism. The steps are quite similar to
those of Sections~\ref{sad}, \ref{loop1}, and \ref{loop2}, and we confine
ourselves to indicating points of difference. We define the operator
\begin{equation}
\label{eq6.12}
A^{\dagger} =  \sum_{1 \leq j_1 < j_2 < \ldots < j_k \leq l}
a_{j_1}^{\dagger} \ldots a_{j_k}^{\dagger} \ .
\end{equation}
This operator is obviously different from the one in Section~\ref{low}
where the same symbol was used. After averaging and the
Hubbard--Stratonovich transformation, we obtain the effective
Lagrangean
\begin{equation}
\label{eq6.13}
\frac{1}{2} {\rm tr}_{\mu} {\rm trg} \sigma^2 + {\rm tr}_{\mu} {\rm
  trg} \ln \biggl[ ( E - \langle \mu | A^{\dagger} \sigma A | \mu
  \rangle - J - \frac{1}{2} \varepsilon L ) \delta_{\mu \nu} \biggr] \
  . 
\end{equation}
The quantity $\sigma$ is both, a supermatrix and a matrix in the
Hilbert space with $m - k$ particles. The saddle--point equation for
$\sigma$ can be solved as in Section~\ref{sad} and yields the solution
$\langle \nu | A^{\dagger} \sigma^{(0)} A | \nu \rangle =
\tau^{(0)}_{\max}$ which is independent of $\nu$. The quantity
$\tau^{(0)}_{\max}$ has a form similar to that of $\tau^{(0)}$ in
Eq.~(\ref{eq6.6}) and is given by $\tau^{(0)}_{\rm max} = E/2 \pm i
\sqrt{N(k) - (E/2)^2}$, with $N(k) = (1/N) \ {\rm tr}_{\mu}
(A^{\dagger} A A^{\dagger} A)$ a constant and $N(k) \gg 1$. This shows
that the average spectrum has semicircle shape with radius $2
\sqrt{N(k)}$. The saddle--point manifold is again generated by the
matrices $T$. Expanding the effective Lagrangean to first order in
$\epsilon$ leads to a term of the form~(\ref{eq6.6a}). This shows that
the level fluctuations are of Wigner--Dyson type. All of this is quite
straightforward. The crucial question is: Do the leading terms in the
loop expansion vanish for $l \rightarrow \infty$?

We consider the one--point function and write $\sigma = \sigma^{(0)} +
\delta \sigma$. The logarithmic term contains the matrix elements
$\langle \mu | A^{\dagger} \delta \sigma A | \mu \rangle$ of $\delta
\sigma$. The steps are the same as in Section~\ref{loop1} but can be
carried through independently for each value of $\mu$, cf.
Appendix~\ref{loop}.  The correction is proportional to $({m \choose
  k}/N(k))^4$. Even in the worst case $k = 1$, this contribution
vanishes like $l^{-4}$ for $l \rightarrow \infty$. For the two--point
function, we find likewise that the first non--vanishing term in the
loop expansion behaves like $({m \choose k}/(N(k))^4$. We conclude
that EGUE$_{\rm max}(k)$ has the same spectral properties as GUE. This
is consistent with results by Mirlin and Fyodorov~\cite{mir91} who
showed that sparse random matrices possess Wigner--Dyson statistics as
long as the number of non--zero matrix elements per row and column
exceeds the value unity.

\section{The Case $k \ll m \ll l$} 
\label{bin}

The results of the previous Sections show that for $2k > m$,
EGOE($k$) and EGUE($k$) behave generically: The average spectrum has
semicircle shape, and the spectral fluctuations coincide with those of
GOE and GUE, respectively. We now address the question: What happens
in the regime $2k \leq m$? We expect a transition to Gaussian spectral
shape but have little information yet on the spectral fluctuation
properties. In this situation, it may be useful to consider a case as
far removed as possible from the transition point $2k = m$, i.e., to
consider $k \ll m$. For technical reasons, it is then advantageous to
take also $m \ll l$. This is the case dealt with in the present
Section.

As a preparatory step, we consider in Section~\ref{bin1} the case $1 =
k \ll m \ll l$. We show that the spectral fluctuations are Poissonian.
By calculating the inverse participation ratio, we show, furthermore,
that the eigenstates display localization in Fock space. Both features
are non--generic. It is conceivable, of course, that the case $k = 1$
is special, and that already $k = 2$ behaves very differently. In
Section~\ref{bin2}, we show that this is not the case. We use the
``binary correlation appoximation'' developed by French and
collaborators~\cite{mon75,bro81,kot00}. With the help of this
approximation, it was shown~\cite{mon75} that for $k \ll m \ll l$, the
average spectrum has Gaussian shape. We generalize the approximation
in such a way that it also yields information on two-- and higher
$n$--point correlation functions. We show that the level fluctuations
are Poissonian for $k \ll m \ll l$.

\subsection{The Case $k = 1$}
\label{bin1}

To simplify notation we again confine ourselves to the unitary case.
The extension to the orthogonal case is perfectly straightforward. For
$m = 1$ the random one--body operator can be diagonalized for each
realization of the ensemble. With $\alpha = 1,\ldots,l$, we denote the
eigenvalues and eigenfunctions by $\varepsilon_{\alpha}$ and
$\Psi_{\alpha}$, respectively. The eigenfunctions can be expanded in
our single--particle basis, $\Psi_{\alpha} = c_{\alpha}^{\dagger} | 0
\rangle = \sum_{j=1}^l U_{\alpha j} a_j^{\dagger} | 0 \rangle$. For $l
\rightarrow \infty$, the eigenvalues $\varepsilon_{\alpha}$ obey GUE
statistics, the average level density has semicircle shape, and the
elements of the matrix $U_{\alpha j}$ are complex Gaussian distributed
random variables with mean value zero and a second moment given by
\begin{equation}
\label{eq7.5}
\overline{U_{\alpha i} U^*_{\beta j}} = \frac{1}{l} \ \delta_{\alpha
  \beta} \delta_{i j} \ .
\end{equation}
For terms of higher order, we use the results of Ullah and Porter
given in Eq.~(7.16) of Ref.~\cite{bro81}.

For $m > 1$, we can easily find the eigenvalues and eigenfunctions of
this ensemble. The eigenfunctions are orthonormal Slater determinants
given by $| \chi_n \rangle = c_{\alpha_1}^{\dagger} \ldots
c_{\alpha_m}^{\dagger} | 0 \rangle$ where the $\alpha$'s take values
from $1$ to $l$ with $\alpha_1 < \ldots < \alpha_m$. The corresponding
eigenvalues are
\begin{equation}
\label{eq7.1}
E_n = \sum_{j=1}^m \varepsilon_{\alpha_j} \ .
\end{equation}
It is obvious that for $m \gg 1$, the average level density has
Gaussian shape: It is the convolution of $m$ semicircular level
densities. It is equally obvious that for $l \gg m \gg 1$, the
spectral statistics of EGUE(1) is Poissonian. Indeed, from
Eq.~(\ref{eq7.1}) we see that neighboring eigenvalues $E_n$ and
$E_{n+1}$ may be composed of quite different combinations of
single--particle energies $\varepsilon_j$, and both level repulsion
and stiffness of the single--particle spectrum lose importance. This
is, to some extent, true already for $m = 2$. 

The example is instructive because for $m = 2$, we are right at the
transition point $2k = m$ separating the regime of the semicircle from
that of the Gaussian, see Sections~\ref{loop1} and \ref{loop2}. The
correction~(\ref{eq6.11}) derived in Section~\ref{loop2} contains the
factor $R$ which vanishes asymptotically for $l \rightarrow \infty$
whereas the case $m = 2$ discussed above leads us to expect that
non--vanishing finite corrections to the Wigner--Dyson fluctuation
behavior exist in this limit. This fact reconfirms our expectation
that higher--order terms in the loop expansion yield contributions of
the form $\bigl( {\rm trg} L T L T^{-1} \bigr)^2$ which do not vanish
asymptotically.

We turn to the eigenfunctions and ask whether these display
localization in Fock space. This would be another strong indication
for non--generic behavior. To this end, we expand the eigenfunctions
in terms of the Slater determinants $| \mu \rangle$ introduced in
Section~\ref{defa},
\begin{equation}
\label{eq7.2}
| \chi_n \rangle = \sum_{\mu=1}^N g_{n \mu} | \mu \rangle
\end{equation}
and calculate the inverse participation ratio
\begin{equation}
\label{eq7.3}
{\cal P} = N \sum_{\mu=1}^N \overline{|g_{n \mu}|^4} \ .
\end{equation}
For states which are completely mixed, we expect $\overline{|g_{n
    \mu}|^2} \cong 1/N$ thus $\overline{|g_{n \mu}|^4} \cong 1/N^2$
and ${\cal P} \approx 1$. A value of ${\cal P}$ significantly larger
than unity signals localization.

The eigenvector $| \chi_n \rangle$ has $m$ occupied single--particle
states labelled $\alpha_p$ with $p = 1,\ldots,m$ and $\alpha_1 <
\ldots < \alpha_m$. For $| \mu \rangle$, the corresponding indices are
$j_q$. The expansion coefficients $g_{n \mu} = \det (U_{\alpha_p j_q})$
are determinants formed of the matrix elements of the unitary
transformation $U_{\alpha j}$ introduced above. It is straightforward
to show that $|g_{n \mu}|^4$ can be written as a determinant,
\begin{equation}
\label{eq7.4}
|g_{n \mu}|^4 = \det \ \biggl( \sum_{{i, j = j_1,\ldots,j_m} \atop
  {\beta = \alpha_1,\ldots,\alpha_m}} U_{\alpha_p i} U^*_{\beta i}
  U_{\beta j} U^*_{\alpha'_p j} \biggr) \ .
\end{equation}
In averaging $|g_{n \mu}|^4$ over the ensemble, we keep the leading
terms in an expansion in inverse powers of $m$. The average is
calculated using Eq.~(\ref{eq7.5}) and Wick contraction. The leading
contribution in powers of $1/m$ is obtained by contracting pairs of
$U$'s in such a way that the minimum number of summations over the
indices $i$ and $\beta$ is destroyed by the Kronecker delta's on the
right--hand side of Eq.~(\ref{eq7.5}). Consider an arbitrary entry in
the matrix appearing in the determinant in Eq.~(\ref{eq7.4}). The
indices $\beta$ and $i$ of the second factor $U^*_{\beta i}$ are both
summed over, suggesting that this factor should be contracted with
either of its neighbors to the right and left. Doing so and using
Eq.~(\ref{eq7.5}) we find for the entry the value $2 (m/l) \sum_i
U_{\alpha_p i} U^*_{\alpha'_p i}$. The factor two accounts for both
ways of performing the contraction. Ullah--Porter corrections to this 
result are small of order $m/l$. Hence,
\begin{equation}
\label{eq7.6}
\overline{|g_{n \mu}|^4} \approx \left( \frac{2 m}{l} \right)^m \
  \overline{\det \sum_i \ U_{\alpha_p i} U^*_{\alpha'_p i}} \ .
\end{equation}
Inserting this result back into the expression~(\ref{eq7.3}) for
${\cal P}$ and using normalization, we obtain ${\cal P} \approx N
(\frac{2 m}{l})^m$. We take the limit $l \rightarrow \infty$ of ${\cal
  P}$ for fixed $m \gg 1$ and find ${\cal P} \rightarrow (2e)^m \gg
1$. We recall that we expect ${\cal P} \approx 1$ for delocalized
states and conclude that for $k \ll m \ll l$, the eigenstates of
EGUE($k$) show localization in Fock space. This statement applies
uniformly to all eigenstates and is not connected with a
delocalization transition.

\subsection{Binary Correlation Approximation}
\label{bin2}

The results of Section~\ref{bin1} raise the question: are these
results specific for the case $k = 1$ (which then would play a
somewhat singular role), or are they typical for $k \ll m \ll l$?
With the help of a suitable generalization of the binary correlation
approximation~\cite{mon75,bro81,kot00}, we now show that the second
alternative holds. We proof that the eigenvalues of the EGUE($k$) have
a Poissonian distribution in the limit $k \ll m \ll l$ in general.

The binary correlation method has been successfully applied to
calculate the average one--point function of the
EGUE($k$)~\cite{mon75,ver84}. In Ref.~\cite{ver84}, the average
one--point function was expanded in a power series in $V_k$. One makes
use of the fact that $V_k$ is a Gaussian distributed random operator
and performs Wick contractions to carry out the average. In the limit
$k \ll m \ll l$, each pair of Wick--contracted $V_k$'s may be
approximately replaced by $v_0^2 \Lambda^{(0)}(k)$. The resulting
expression can be resummed using Borel summation. This yields the
Gaussian spectral shape first found by Mon and French~\cite{mon75} .

We proceed similarly for the two--point function by expanding both
Green functions in powers of $V_k$ and applying Wick contraction. The
disconnected part is calculated in exact analogy to Ref.~\cite{ver84}.
However, cross--contracted pairs of $V_k$'s which arise in the
connected part have to receive special treatment. In
Appendix~\ref{binary}, we show that in the terms of order $(V_k)^{2n}$,
the contributions of cross--contracted pairs carry lower powers of $l$
than the contributions to the disconnected part. In the limit $k \ll m
\ll l$, this fact causes the connected part to vanish relative to the
disconnected part. More precisely, the properly normalized two--point
correlation function
\begin{equation}
R_2(z_1,z_2) = \frac{\overline{g(z_1) g(z_2)}}{\overline{g(z_1)} \cdot
  \overline{g(z_2)}} - 1
\label{eq7.7}
\end{equation}
vanishes in the limit $l \rightarrow \infty$. More generally, all
higher $n$--point functions turn out to be asymptotically given by the
product of $n$ average one--point functions. Hence, all connected
$n$--point correlation functions vanish in the limit $k \ll m \ll l$,
yielding Poissonian spectral fluctuations. 
  
The binary correlation approximation applies for non--zero values of
$k$ as long as $k \ll m \ll l$. The approximation predicts Gaussian
spectral shape and, perhaps unexpectedly, Poissonian spectral
fluctuations. This is in keeping with the results trivially expected
for $k = 0$ and with the results for $k = 1$ found in
Section~\ref{bin1}. In terms of spectral properties, the cases $k=0$
and $k=1$ are seen not to play a special role. In the limit $k \ll m
\ll l$, the spectral properties do not show discontinuous changes when
$k$ approaches zero. This corresponds to the continuous dependence on
$k$ in all the formulae derived in earlier Sections.

\section{Conclusions} 
\label{con}

We have studied the shape of the average spectrum and the eigenvalue
fluctuations of the embedded ensembles EGOE($k$) and EGUE($k$) in the
limit of infinite matrix dimension, attained by letting the number $l$
of degenerate single--particle states go to infinity. We have shown
that for sufficiently high rank $k$ of the random interaction ($2k >
m$ where $m$ is the number of Fermions), these ensembles behave
generically: The spectrum has semicircle shape, and the eigenvalue
fluctuations obey Wigner--Dyson statistics. A smooth transition to a
different regime takes place at or near $2k = m$. It has long been
known that the average spectrum changes into Gaussian shape, although
the point of departure from the semicircle shape was not known. We
have shown that in addition --- and contrary to general expectations
--- the level fluctuations also change and are not of Wigner--Dyson
type for $2k \lesssim m$. We cannot pin down precisely the $k$--value
where such change occurs. The case $m = 2, k = 1$ suggests that this,
too, happens near $2k = m$. We have proved, however, that the level
fluctuations become Poissonian in the limit $k \ll m \ll l$. Our
analysis of the case $k = 1$ suggests that in this limit the
eigenfunctions display localization in Fock space. With the analytical
methods at our disposal, we cannot reach that regime of the embedded
ensembles where neither Poissonian nor Wigner--Dyson statistics
applies. This is no surprise: We do not know of any case where such an
aim would have been achieved analytically.

It is interesting to compare our results with those of
Ref.~\cite{ver84}. There the two--point correlation function was
calculated to lowest order in the loop expansion. (This expansion
differs from the one in the present paper). It was found that the
two--point function did have the dependence on $r$ expected from
canonical random matrix theory. Here, $r$ is the difference in energy
of the two Green functions in units of the mean level
spacing. However, the coefficient multiplying $r^{-2}$ turned out to
be too large. It was speculated that higher--order terms in the loop
expansion would reduce this coefficient to the value unity expected
from canonical RMT. Now we know that this speculation applies only for
$2k > m$ or so. In the limit $k \ll m \ll l$, the higher--order terms
in the loop expansion of Ref.~\cite{ver84} must actually cancel the
contribution of lowest order.

Within the framework of the supersymmetry approach and in diagrammatic
language, the transition at $2k = m$ is caused by the increased weight
(with decreasing $k$) of intersecting Wick contraction lines. Universal
random--matrix results are obtained whenever such contributions are
negligible, and this apparently is the case for $2k > m$. In other
words, the intersecting contraction lines are responsible for the
deviations from random--matrix universality. We ascribe the special
role played by the value $2k = m$ of the rank $k$ to duality. The
duality relation plays an important role in obtaining our results. It
connects $k$ with $(m-k)$ and, thus, assigns a special role to the
value $2k = m$.

From a physical point of view, our results can be understood in terms
of the ratio $K_{\beta}/P$ of the number of uncorrelated random
variables over the number of independent links, i.e., the number
of non--zero entries in the matrix representation of the $k$--body
interaction in Hilbert space. We have shown that if all links were
to carry uncorrelated random variables, the ensembles would have
Wigner--Dyson type spectral fluctuations. Conversely, if all links
were to carry the same random variable, the ensembles would be
completely integrable and display Poissonian statistics. These
statements hold for all values of $k$. The actual situation is located
between these two limits. The case $k = 1$ is closest to the
integrable case, and the case $k = m$ corresponds to canonical
random--matrix theory. These facts show that deviations from
random--matrix universality are not caused by the number of zeros in
the matrix representation of the interaction but are strictly due to
correlations between matrix elements. We may think of the random
$k$--body interaction as a sum of terms, each term carrying a single
random variable multiplied by a fixed matrix. All random variables in
the sum are uncorrelated. Reducing the number of independent variables
is tantamount to regrouping this sum into a sum with a smaller number
of terms. The question whether complete mixing between all states in
Hilbert space is attained and random--matrix type results are
obtained, is then a question of the number of terms in the sum. For
instance, a stochastic coupling of only two fixed matrices will, in
general, not lead to random--matrix results.

In the light of these remarks, it seems appropriate to use the term
localization for the results on eigenfunctions derived for $k = 1$.
Indeed, Anderson localization may be viewed as a problem of insufficient
mixing in a stochastic Hamiltonian. This insufficient mixing is due to
the restriction of the interaction to nearest--neighbor couplings. If
each lattice site were coupled to all other sites, localization would
not occur. In the present context, it is not the limited number of
links which causes localization but the further constraint on
stochasticity caused by correlations between the matrix elements on
these links.

\begin{figure}
\begin{minipage}{15cm}
\centerline{\psfig{figure=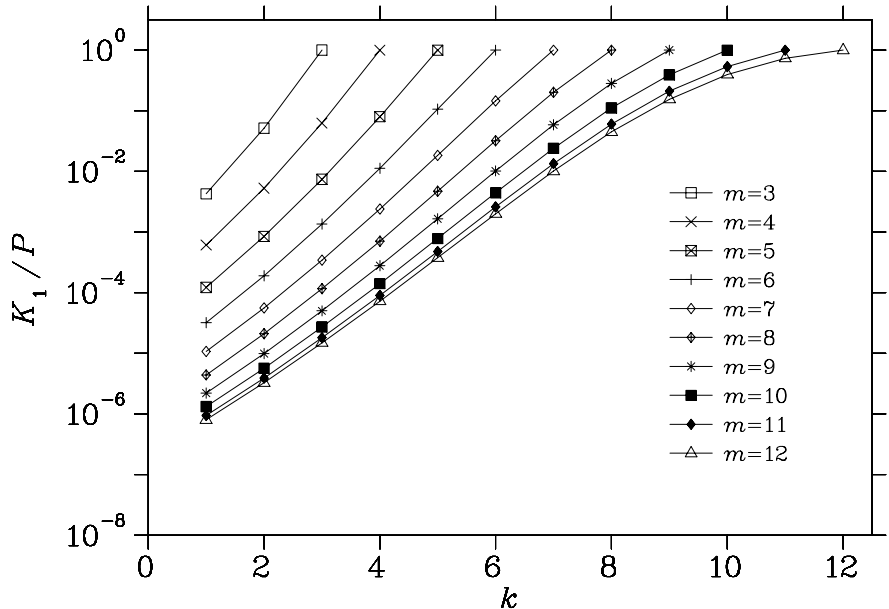,width=8cm,angle=90}}
\end{minipage}
\caption{The ratio $K_1/P$ on a logarithmic scale versus k for $l =
  25$ and for various values of $m$.}
\label{fig3}
\end{figure}

These results shed new light on the agreement of numerical results for
EGOE(2) and EGUE(2), and of experimental data obtained in nuclear and
atomic physics, with Wigner--Dyson spectral statistics. (We disregard
here the complications which arise because the single--particle levels
in major shells are non--degenerate in atoms and nuclei). Indeed, it is
true both for the numerical work and for the experimental data that
the underlying spaces are very far from the limit $l \rightarrow
\infty$ considered in this paper. Among all these cases, the case of
nuclei with excitation energies near neutron threshold is probably the
one with the largest number of interacting configurations: The number
of valence nucleons is around six or eight, and the total number of
configurations is around $10^6$, implying values of $l$ in the tens or
twenties. A comparison of Figure~\ref{fig3} with Figure~\ref{fig1}
shows that in this situation, the ratio $K_1 / P \approx 10^{-4}$
or so is relatively large. Thus, the fluctuations are near the GOE or
GUE limit although for $l \rightarrow \infty$, none of these systems
would display universal random--matrix behavior. Putting things very
pointedly, we might say that the agreement between Wigner--Dyson level
statistics and the results found in these systems, is a finite--size
effect.

Acknowledgment. We are grateful to O. Bohigas and T.H. Seligman. Both
helped us with many questions and suggestions and were actively
involved in some aspects of this work. We thank T. Papenbrock for a 
careful reading of the manuscript.

\appendix

\section{Calculation of Loop Corrections}
\label{loop}

\subsection{Loop Corrections for the EGUE($k$)}

The loop corrections are obtained in terms of an expansion around the
saddle--point solution. The limit $l \to \infty$ is taken. For each
value of $(s,a)$, let $\delta\sigma^{(sa)}$ denote the deviation of
$\sigma^{(s a)}$  from the saddle--point value. First, we expand the
logarithmic term in the effective Lagrangean in powers of $\delta
\Sigma = \sum_{sa} \lambda^{(s)}(k) C^{(sa)} \delta \sigma^{(sa)}$ and
in powers of the source variable $J$. The resulting terms are denoted
by $(p,q)$ where $p$ indicates the power of $\delta\Sigma$ and $q$ the
power of $J$. Second, the resulting exponential is expanded again in
powers of $\delta\Sigma$ and $J$. We keep terms up to fourth order in
$\delta\Sigma$ and up to first (second) order in $J$ for the one-- 
(two)--point function, respectively. Finally, we perform the Gaussian
integration over $\delta\sigma^{(sa)}$ and identify among the
remaining terms those which are of leading order in the limit
$l\rightarrow\infty$.

{\it One--point function}: After expanding the logarithmic term, we retain
the terms $(p,q)$ with $p=1,\ldots,4$ and $q=0,1$. The term $(0,0)$
vanishes identically. Because of the saddle--point condition 
Eq.~(\ref{eq6.3}), the term $(1,0)$ cancels the corresponding term
from the expansion of $(N/2) \sum_{sa} {\rm trg}[\sigma^{(sa)}]^2$.
After taking the trace over $\mu$, we combine the term $(2,0)$ with
the quadratic term from the expansion of $(N/2) \sum_{sa} {\rm
  trg}[\sigma^{(sa)}]^2$ and obtain a new Gaussian exponent of the
form $- (N/2)$ $\sum_{sa} [1 - (\tau^{(0)})^2$ $(\Lambda^{(s)}(k) /
\Lambda^{(0)}(k))]$ ${\rm trg}$ $[\delta \sigma^{(sa)}]^2$. We are
left with the terms $(3,0)$, $(4,0)$ and with the terms $(p,1)$ for
all $p=0,\ldots,4$ as listed in Table~\ref{tab_pq_one}. Expanding the
exponential in powers of these terms, we keep terms up to fourth order
in $\delta\sigma^{(sa)}$ but omit terms odd in $\delta\sigma^{(sa)}$
as these vanish upon the Gaussian integration. Moreover, we keep only
terms linear in $J$ and obtain $[(0,1) + (2,1) + (4,1) + (3,0)(1,1) +
(4,0)(0,1)]$. The term $(4,0)(0,1)$ vanishes due to the Efetov--Wegner
theorem. Evaluation of the remaining terms with the help of the
orthogonality relations~(\ref{eq3.11}), the duality
relation~(\ref{eq3.18}), and Wick's theorem for the Gaussian integrals
is straightforward and yields non--zero contributions only from
$(0,1)$ and from $(4,1)$. This gives the result in Eq.~(\ref{eq6.7}).

\begin{table}[t!]
\begin{center}
\begin{tabular}{c c l}
\hline
\hline\\
\hspace{3mm}$p$\hspace{3mm} & \hspace{6mm}$q$\hspace{6mm} & \\ \\
\hline\\
$3$ & $0$ & $-\frac{1}{3}\left(\frac{\tau^{(0)}}{\lambda^{(0)}}\right)^3\
             {\rm tr}_\mu {\rm trg} (\delta\Sigma)^3$\\ \\
$4$ & $0$ & $-\frac{1}{4}\left(\frac{\tau^{(0)}}{\lambda^{(0)}}\right)^4\
             {\rm tr}_\mu {\rm trg} (\delta\Sigma)^4$\\ \\
$0$ & $1$ & $-\frac{\tau^{(0)}}{\lambda^{(0)}}\ {\rm tr}_\mu {\rm trg} J$\\ \\
$1$ & $1$ & $-\left(\frac{\tau^{(0)}}{\lambda^{(0)}}\right)^2\
             {\rm tr}_\mu {\rm trg} (\delta\Sigma J)$\\ \\
$2$ & $1$ & $-\left(\frac{\tau^{(0)}}{\lambda^{(0)}}\right)^3\
             {\rm tr}_\mu {\rm trg} \left((\delta\Sigma)^2 J\right)$\\ \\
$3$ & $1$ & $-\left(\frac{\tau^{(0)}}{\lambda^{(0)}}\right)^4\
             {\rm tr}_\mu {\rm trg} \left((\delta\Sigma)^3 J\right)$\\ \\
$4$ & $1$ & $-\left(\frac{\tau^{(0)}}{\lambda^{(0)}}\right)^5\
             {\rm tr}_\mu {\rm trg} \left((\delta\Sigma)^4 J\right)$\\ \\
\hline
\hline
\end{tabular}
\caption{Relevant terms in the expansion of the logarithmic term in
  the effective Lagrangean of the EGUE($k$) one--point function.
  Parameters $p$ and $q$ indicate powers of $\delta\Sigma$ and $J$,
  respectively.\label{tab_pq_one}}
\end{center}
\end{table}

{\it Two--point function}: After expanding the logarithm, we retain all
terms $(p,q)$ with $p=1,\ldots,4$ and $q=0,1,2$. Again, $(0,0)$
vanishes identically, $(1,0)$ cancels and $(2,0)$ forms part of the
new Gaussian  exponent. For convenience, we keep the terms $(0,1)$ and
$(0,2)$ in the exponent as well. We are left with the terms $(3,0)$,
$(4,0)$, $(p,1)$ and $(p,2)$ for $p=1,\ldots,4$ as shown in
Table~\ref{tab_pq_two}. The exponential is expanded in a power series
where we keep again only terms even and up to fourth order in
$\delta\Sigma$, and terms up to second order in the source variable
$J$. Performing the Gaussian integration and dropping terms which
vanish as $1/N$ for $N \rightarrow \infty$, we are finally left only
with the contributions stemming from $(4,1)$ and $(2,1)(2,1)$. We
rewrite these terms as an exponential. From $(4,1)$, we obtain a
correction to the canonical random--matrix contribution as given in
expression~(\ref{eq6.10}). The only non--trivial correction stems from
$(2,1)(2,1)$ and is proportional to the expression~(\ref{eq6.11}).

\begin{table}[t!]
\begin{center}
\begin{tabular}{c c l}
\hline
\hline\\
\hspace{3mm}$p$\hspace{3mm} & \hspace{6mm}$q$\hspace{6mm} & \\ \\
\hline\\
$3$ & $0$ & $\frac{i}{3}(\lambda^{(0)})^{-3}\
            {\rm tr}_\mu {\rm trg} (L\ \delta\Sigma)^3$\\ \\
$4$ & $0$ & $\frac{1}{4}(\lambda^{(0)})^{-4}\
            {\rm tr}_\mu {\rm trg} (L\ \delta\Sigma)^4$\\ \\
$1$ & $1$ & $           (\lambda^{(0)})^{-2}\
            {\rm tr}_\mu {\rm trg} (L\ \delta\Sigma\ L\ \tilde{J})$\\ \\
$2$ & $1$ & $          i(\lambda^{(0)})^{-3}\
            {\rm tr}_\mu {\rm trg} \left((L\ \delta\Sigma)^2\ L\
              \tilde{J}\right)$\\ \\
$3$ & $1$ & $          -(\lambda^{(0)})^{-4}\
            {\rm tr}_\mu {\rm trg} \left((L\ \delta\Sigma)^3\ L\
              \tilde{J}\right)$\\ \\
$4$ & $1$ & $           -i(\lambda^{(0)})^{-5}\
            {\rm tr}_\mu {\rm trg} \left((L\ \delta\Sigma)^4\ L\
              \tilde{J}\right)$\\ \\
%$0$ & $2$ & $\frac{1}{2}\left(\frac{\tau^{(0)}}{\lambda^{(0)}}\right)^2
%            {\rm tr}_\mu {\rm trg} (L\ \tilde{J})^2$ \\ \\
$1$ & $2$ & $            i(\lambda^{(0)})^{-3}\
            {\rm tr}_\mu {\rm trg} \left(L\ \delta\Sigma\ (L\
              \tilde{J})^2\right)$\\ \\
$2$ & $2$ & $            -(\lambda^{(0)})^{-4}\
      {\rm tr}_\mu {\rm trg} \left[(L\ \delta\Sigma)^2 (L\ \tilde{J})^2
      + \frac{1}{2} (L\ \delta\Sigma\ L\ \tilde{J})^2\right]$\\ \\
$3$ & $2$ & $           -i(\lambda^{(0)})^{-5}\
      {\rm tr}_\mu {\rm trg} \left[(L\ \delta\Sigma)^3 (L\ \tilde{J})^2
      + (L\ \delta\Sigma\ L\ \tilde{J})^2\right]$\\ \\
$4$ & $2$ & $             (\lambda^{(0)})^{-6}\
      {\rm tr}_\mu {\rm trg} \left[(L\ \delta\Sigma)^4 (L\ \tilde{J})^2
      + (L\ \delta\Sigma)^2 (L\ \delta\Sigma\ L\ \tilde{J})^2\right.$\\
    &     & \hspace{40mm}$\left.+ \frac{1}{2} \left((L\ \delta\Sigma)^2\ L\
            \tilde{J}\right)^2\right]$\\  
    \\
\hline
\hline
\end{tabular}
\caption{Relevant terms in the expansion of the logarithmic term in
  the effective Lagrangean of the EGUE($k$) two--point function.
  Parameters $p$ and $q$ indicate powers of $\delta\Sigma$ and $J$,
  respectively. The expressions are evaluated at the center of the
  spectrum, $E=0$. We put $\tilde{J} = (\epsilon/2) TLT^{-1} +
  TJT^{-1}$.\label{tab_pq_two}}
\end{center}
\end{table}

\subsection{Loop Corrections for the Limiting Ensemble\\ EGUE$_{\rm
    max}(k)$}

For the limiting ensemble EGUE$_{\rm max}(k)$, the loop corrections
are calculated in very much the same way as for the EGUE($k$). It is
important, however, to keep in mind the operator character of $\sigma$
in the Hilbert space of the $(m-k)$--particle system. The logarithmic
term in the effective Lagrangean contains the matrix elements
$\langle\mu|A^\dag\delta\sigma A|\mu\rangle$ of $\delta\sigma$. The
calculation is carried through independently for each value of $\mu$.
After expansion of the logarithm, it is convenient to include in the
expansion of the exponent also the term $(2,0)$. We again restrict
ourselves to terms up to fourth order in $\delta\sigma$ and up to
first or second order in $J$.

{\it One--point function}: After expansion of the exponential, we are left
with $[(0,1)+(2,1)+(4,1)+(2,0)(0,1)+(2,0)(2,1)+(3,0)(1,1)+(4,0)(0,1)]$.
The terms $(2,0)(0,1)$ and $(4,0)(0,1)$ yield zero due to the
Efetov--Wegner theorem. After the Gaussian integration we are left
with contributions from the canonical term $(0,1)$ and from $(4,1)$.
Both together yield
\begin{equation}
-\frac{N\tau^{(0)}_{\rm max}}{N(k)} {\rm trg} J \left[1+\left(\frac{{m 
        \choose k}}{N(k)}\right)^4\right]\,.
\end{equation}
The correction to the leading term is $({m \choose k}/N(k))^4$. Even
in the worst case, $k=1$, this contribution vanishes like $l^{-4}$ for
$l\rightarrow\infty$.

{\it Two--point function}: After expanding the exponential and performing
the Gaussian integration, we are left with the terms
$[(0,2)+(2,2)+(4,2)+(2,0)(2,2)+(0,1)(0,1)+(0,1)(4,1)+
(1,1)(1,1)+(1,1)(3,1)+(2,1)(2,1)+(2,0)(1,1)(1,1)]$. Except for the
canonical contribution $(0,1)(0,1)$, all these terms vanish at least 
like $({m \choose k}/N(k))^4$, as does the correction term for the
one--point function.

\section{Binary Correlation Approximation for the $n$--Point Funtion}
\label{binary}

In this Appendix, we show that for $k \ll m \ll l$, the eigenvalues of
EGUE($k$) have a Poissonian distribution. The proof applies
analogously to EGOE($k$). Our proof extends the calculation of the
average one--point function of Section 5.2 of Ref.~\cite{ver84} to the
case of the $n$--point function with $n > 1$. In order to avoid the
occurrence of terms which diverge in the limit $l \rightarrow \infty$
in the sums introduced below, we choose in this Appendix the second
moment of the matrix elements of $V_k$ differently from the main body
of the paper (cf. Eq.~(\ref{eq2.4})) by putting $v_0^2 =
(\Lambda^{(0)}(k))^{-1}$. This choice has the merit of yielding an
average Gaussian spectrum with unit width for all choices of $k,m$,
and $l$.

We briefly recall the method of Ref.~\cite{ver84}. The average
one--point function
\begin{equation}
\overline{g(z)} = \frac{1}{N} {\rm tr}_{\mu} \biggl (
\overline{\frac{1}{z - V_k}} \biggr )
\label{eq1}
\end{equation}
is expanded in inverse powers of $z$ and the summation and the average
over the ensemble are interchanged,
\begin{equation}
\overline{g(z)} = \frac{1}{N} \sum_{p = 0}^{\infty} \frac{1}{z^{2p+1}}
  {\rm tr}_{\mu} \biggl ( \overline{(V_k)^{2p}} \biggr ) \ .
\label{eq2}
\end{equation}
Because of the Gaussian distribution of $V_k$, only even powers of
$V_k$ contribute. The average is performed via Wick contraction. There
are $(2p-1)!!$ ways of pairwise contracting the $V_k$'s. For $k \ll m
\ll l$, each contracted pair of $V_k$'s contributes a factor unity.
The result is 
\begin{equation}
\overline{g(z)} = \frac{1}{N} \sum_{p = 0}^{\infty} \frac{1}{z^{2p+1}}
(2p-1)!! \ .
\label{eq3}
\end{equation}
Borel summation of this series gives a closed expression for
$\overline{g(z)}$ and the Gaussian form for the average level density,
see Ref.~\cite{ver84}.

In considering next the average two--point function $\overline{g(z_1)
  g(z_2)}$, we use the expansion~(\ref{eq2}) for $g(z_1)$ and for
$g(z_2)$. With $z_1 = E^+ + \varepsilon/2$ and $z_2 = E^- -
  \varepsilon/2$, the letters $E$ and $\varepsilon$ denote the mean
value and the difference of the energy arguments of the two Green
functions, respectively. An upper plus (minus) sign denotes an
infinitesimal positive (negative) imaginary energy increment. We are
interested in values of $\varepsilon$ which are of the order of the
mean level spacing and, thus, small compared to $|E|$ with $|E| \gg 1$.
(We recall that the spectrum has width unity). Then, we have
approximately $|z_1| \sim |z_2|$. In the expansion, we collect all
terms which carry the same even power $2n$ of $V_k$. This yields
\begin{equation}
\overline{g(z_1) g(z_2)} = \sum_{n = 0}^{\infty} \sum_{t = 0}^{2n}
\frac{1}{z_1^{t+1} z_2^{2n+1-t}} \frac{1}{N^2} \overline{{\rm tr}_{\mu}
  [(V_k)^t] \ {\rm tr}_{\nu} [(V_k)^{2n-t}]} \ .  
\label{eq4}
\end{equation}
Wick contraction may lead to $s = 0,1,\ldots,n$ cross--contracted
pairs of $V_k$'s, i.e., pairs where the two $V_k$'s reside in different
traces. The remaining number of $V_k$'s in each trace must be even.
Hence,
\begin{eqnarray}
\overline{g(z_1) g(z_2)} &=& \sum_{n = 0}^{\infty} \sum_{s = 0}^{n}
\sum_{p=0}^{n-s} \frac{1}{z_1^{s+2p+1} z_2^{2n+1-s-2p}}\nonumber\\
&& \qquad\quad \times \ \frac{1}{N^2}
\biggl ( \overline{{\rm tr}_{\mu} [(V_k)^{s+2p}] \ {\rm tr}_{\nu}
  [(V_k)^{2n-s-2p}]} \biggr )_s \ .  
\label{eq5}
\end{eqnarray}
The index $s$ indicates that $s$ pairs are cross--contracted. The
number of different ways of cross--contracting the $V_k$'s is ${s+2p
  \choose s} {2n-2p-s \choose s}$, while there are $(2p-1)!! \
(2n-2p-1)!!$ ways of pairwise contracting the remaining $V_k$'s. As in
the case of the one--point function, each of the latter contractions
yields the factor unity, irrespective of the sequence in which the
$V_k$'s appear under the trace, provided that $k \ll m \ll l$.
Likewise, for $k \ll m \ll l$ each way of cross--contracting the
remaining $s$ pairs of $V_k$'s yields the same answer, irrespective of
where they appear in each of the traces. The result is
\begin{eqnarray}
\overline{g(z_1) g(z_2)} &=& \sum_{n = 0}^{\infty} \sum_{s = 0}^{n}
\sum_{p=0}^{n-s} \frac{{s+2p \choose s} {2n-2p-s \choose s} (2p-1)!!
  (2n-2p-1)!!} {z_1^{s+2p+1} z_2^{2n+1-s-2p}} \nonumber \\
&& \qquad \times \ \frac{1}{N^2} \biggl ( \overline{{\rm tr}_{\mu}
  [(V_k)^{s}] \ {\rm tr}_{\nu} [(V_k)^{s}]} \biggr )_s \ .  
\label{eq6}
\end{eqnarray}

We now show that for each term with fixed $n$ in Eq.~(\ref{eq6}), in
the limit $l~\rightarrow~\infty$ the terms with $s \neq 0$ become
negligibly small in comparison with the terms that have $s = 0$. The
latter correspond to the unlinked contributions. Upon resummation,
these terms yield the product $\overline{g(z_1)} \cdot
\overline{g(z_2)}$. This shows that the two--point function
factorizes, and that the linked correlation function $R_2(z_1,z_2)$
defined in Eq.~(\ref{eq7.7}) vanishes asymptotically in the limit $l
\rightarrow \infty$.

To this end, we observe that the binomial factors and powers of $z_1$
and $z_2$ in Eq.~(\ref{eq6}) do not depend on $l$ or on $m$. Moreover,
we recall that $|z_1| \sim |z_2|$. Hence, the only $l$--dependent part in
the sum in Eq.~(\ref{eq6}) is
\begin{equation}
T_s = \frac{1}{N^2} \biggl ( \overline{{\rm tr}_{\mu} [(V_k)^{s}] \ 
        {\rm tr}_{\nu} [(V_k)^{s}]} \biggr )_s \ . 
\label{eq7}
\end{equation}
Obviously, we have $T_0 = 1$. Thus, $T_0$ and with it all unlinked
contributions are constant and independent of $l$. To prove our claim,
it suffices to show that $T_s$ vanishes at least as $l^{-1}$ for all
$s \geq 1$.

We write $V_k = \sum_{\alpha \beta} v_{\alpha \beta}
A^{\dagger}_{\alpha} A_{\beta}$, with $v_{\alpha \beta}$ and
$A^{\dagger}_{\alpha}$ a short--hand notation for the random variables
and for $a^{\dagger}_{j_1} \ldots a^{\dagger}_{j_k}$ with $j_1 
< j_2 < \ldots < j_k$, respectively, and likewise for $A_{\beta}$. Then,
\begin{equation}
T_s = \frac{1}{N^2 (\Lambda^{(0)}(k))^s} \sum_{\alpha_1,\beta_1,
  \ldots, \alpha_s, \beta_s} {\rm tr}_\mu (A^{\dagger}_{\alpha_1}
  A_{\beta_1} \ldots A^{\dagger}_{\alpha_s} A_{\beta_s}) {\rm tr}_\nu
  (A^{\dagger}_{\beta_1} A_{\alpha_1} \ldots A^{\dagger}_{\beta_s}
  A_{\alpha_s}) \ .
\label{eq8}
\end{equation}
Without loss of generality, the indices $\alpha_1; \alpha_2; \ldots ;
\alpha_s$ can be chosen as $1,\ldots,k; \\ k+1, \ldots, 2k; \ldots;
(s-1)k+1, \ldots, sk$. This yields an overall factor ${l \choose k}^s$.
We find
\begin{eqnarray}
T_s &=& \frac{{l \choose k}^s}{N^2 (\Lambda^{(0)}(k))^s}
  \sum_{\beta_1, \ldots, \beta_s} {\rm tr}_\mu (A^{\dagger}_{1,\ldots,k}
  A_{\beta_1} \ldots A^{\dagger}_{(s-1)k+1,\ldots,sk} A_{\beta_s})
  \nonumber \\
&& \qquad \qquad \quad \times \ {\rm tr}_\nu (A^{\dagger}_{\beta_1}
A_{1,\ldots,k} \ldots 
  A^{\dagger}_{\beta_s} A_{(s-1)k+1,\ldots,sk}) \ .
\label{eq9}
\end{eqnarray}
The summation over the indices $\beta_1, \ldots, \beta_s$ is now
restricted to a sum over all permutations of the integers
$1,2,\ldots,sk$. We note that the number $(sk)!$ of such permutations
does not depend on $l$ or $m$. We consider one such permutation and
show that the resulting dependence on $l$ is the same for all
permutations.

In both traces, any permutation generates terms of the type
$a^{\dagger}_j  a_j$ and terms of the type $a_j  a^{\dagger}_j$. Let
the number of terms of the first type in the first trace be $i$, those
of the second type in the first trace be $sk-i$. It is easy to see
that the corresponding numbers in the second trace are $sk-i$ and $i$,
respectively. This implies that in the first trace, only those states
$| \mu \rangle$ contribute for which in the first $sk$
single--particle states, $i$ states are occupied. The remaining $m-i$
particles must be distributed over the remaining $l-sk$
single--particle states. This yields a total of ${l-sk \choose m-i}$
Hilbert--space states which contribute to the first trace. The
corresponding counting for the second trace yields ${l-sk \choose
  m-sk+i}$. Altogether, we find that the contribution of this
particular permutation to the quantity $T_s$ is given by
\begin{equation}
\frac{{l \choose k}^s}{N^2 (\Lambda^{(0)}(k))^s} {l-sk \choose m-i} {l-sk
  \choose m-sk+i} \ . 
\label{eq10}
\end{equation}
Using Stirling's formula, we find that for $l \rightarrow \infty$,
this expression behaves as $l^{-sk}$, independent of the value of
$i$. This then shows that all permutations yield the same
behavior. Their number is independent of $l$ and $m$. (Actually, the
single--particle operators appear in different sequences, giving
perhaps rise to cancellations. This possibility is not considered
here). We conclude that all connected terms in the two--point function
vanish at least as fast as $l^{-k}$, so that for $l \rightarrow
\infty$, the two--point function is given by the product of the
averaged one--point functions and the two--point correlation function
$R_2(z_1,z_2)$ in Eq.~(\ref{eq7.7}) vanishes.

The same line of reasoning can be applied to the $n$--point functions
with $n > 2$. We find that every totally connected piece in the
$n$--point function containing $s$ cross--contracted pairs of $V_k$'s
tends to zero as $l^{-sk}$ for $l \rightarrow \infty$. This shows that
all higher $n$--point functions are asymptotically likewise given by
the product of $n$ average Green functions. Thus, all correlation
functions vanish asymptotically, and the spectral fluctuations for
EGOE($k$) become Poissonian in the limit $k \ll m \ll l$.

\end{document}